\documentclass{nsr}

\usepackage{amsmath,graphicx,array}
\usepackage{dcolumn,soul}%

\usepackage{amsthm}
\usepackage[figuresright]{rotating}%
\usepackage{algorithm, algorithmicx, algpseudocode}
\usepackage{listings}%
\usepackage{hyperref}

\newcommand{\kms}{km s$^{-1}$}

\makeatletter
\def\uns{\ifmmode\,\else$\,$\fi}%

\makeatother
 
\jvol{XX}
\jnum{X}
\jyear{Year}
\doi{10.1093/nsr/XXXX}
\received{XX XX Year}
\revised{XX XX Year}
\accepted{XX XX Year}

\begin{document}

\dhead{RESEARCH ARTICLE}

\subhead{EARTH SCIENCES}

\title{High-Frequency Magnetohydrodynamic Waves with Substantial Energy in the Solar Polar Corona }

\author{Yuhang Gao$^{1,2,3}$}

\author{Hui Tian$^{1,2,*}$}

\author{Richard Morton$^4$}

\author{Tom Van Doorsselaere$^{3}$}

\author{Daye Lim$^{5}$}

\author{Mingzhe Guo$^{6,7}$}

\author{Jiansen He$^{1}$}

\author{Zhenyong Hou$^{1,2}$}

\affil{$^1$School of Earth and Space Sciences, Peking University, Beijing, 100871, People's Republic of China}

\affil{$^2$State Key Laboratory of Solar Activity and Space Weather, National Space Science Center, Chinese Academy of Sciences, Beijing 100190, People’s Republic of China}

\affil{$^3$Centre for mathematical Plasma Astrophysics, Department of Mathematics, KU Leuven, Celestijnenlaan 200B bus 2400, B-3001 Leuven, Belgium}

\affil{$^4$Department of Mathematics, Physics and Electrical Engineering, Northumbria University, Newcastle upon Tyne, NE1 8ST, UK}

\affil{$^5$Department of Mathematics and Statistics, University of Exeter, Exeter, EX4 4QF, UK}

\affil{$^6$Institute of Frontier and Interdisciplinary Science, Shandong University, Qingdao, 266237, China}

\affil{$^7$Shandong Key Laboratory of Space Environment and Exploration Technology, Institute of Space Sciences, Shandong University, Shandong, China}

\authornote{\textbf{Corresponding authors.} Email: huitian@pku.edu.cn}


\abstract[ABSTRACT]{The acceleration and heating of the fast solar wind remain long-standing challenges in space physics. One type of leading theoretical models requires high-frequency magnetohydrodynamic (MHD) waves to transport and dissipate sufficient energy in the corona. However, such high-frequency waves with energetically significant amplitudes have never been unambiguously observed, leaving a key gap between theories and observations. Using high-cadence, high-resolution extreme-ultraviolet imaging from Solar Orbiter’s Extreme Ultraviolet Imager, we identify a previously hidden population of high-frequency MHD waves in coronal plumes of the solar polar region. An analysis of the detected propagating kink waves shows that over one-third have periods shorter than 100 s, a population largely undetected by earlier instruments. Power spectral analysis demonstrates that these high-frequency waves carry substantial energy flux, which are significantly underestimated in lower-cadence data. These results suggest that high-frequency MHD waves may contribute importantly to the energy budget of the solar polar corona and could play a role in solar wind acceleration, highlighting the value of high-resolution observations for probing energy transport in magnetized space and astrophysical plasmas.}



\maketitle

\section{Introduction}\label{sec:intro}

The origin and acceleration of fast solar wind remains central challenges in space plasma physics, and serves as one of the main scientific objects of the Solar Polar-orbit Observatory (SPO \cite{SPO2025}), a Chinese solar exploration mission scheduled to be launched in 2029. The solar wind shapes the solar-terrestrial space and drives space weather phenomena which can impact Earth's environment and modern technology \cite{Zhao2011,HeSP2020NSR,Tarduno2025,Shang2025}. One type of leading models for heating and accelerating the solar wind posits that energy is supplied by high-frequency magnetohydrodynamic (MHD) waves, particularly Alfv\'{e}n waves \cite{Marsch1997,He2009}. These waves are expected to efficiently transfer energy from the lower solar atmosphere into the corona and dissipate it locally, thereby contributing significantly to both coronal heating and wind acceleration \cite{tvd2020SSRv,Banerjee2021}.

Despite their remarkable role in theory, a key observational gap remains: high-frequency coronal Alfv\'{e}n waves with energetically significant amplitudes have never been unambiguously detected. Most previous observations \cite{tomczyk2007,McIntosh2011,thurgood2014,morton2015,morton2019,weberg2020,yang2020,yang2020Sci,yang2024} focused on propagating kink waves, transverse oscillations of magnetic flux tubes widely regarded as Alfv\'{e}nic in nature \cite{goossens2009,McIntosh2011}. However, these kink waves predominantly exhibit periods longer than 100 s, with an inferred energy flux typically below 100 W m$^{-2}$ \cite{tomczyk2007,thurgood2014,weberg2018,morton2019}, which is insufficient to account for coronal heating and solar wind acceleration in open-field regions. High-frequency propagating kink waves, which could carry substantially more energy and dissipate it more effectively in the corona \cite{Marsch1997,terradas2010,tvd2020ApJ}, have likely been underdetected due to the limited cadence and spatial resolution of previous instruments. Nevertheless, theoretical models and chromospheric observations suggest that a rich population of high-frequency waves (periods below 100 s) could exist in the corona \cite{He2009,bate2022,kuniyoshi2024}, potentially serving as a major energy reservoir for the solar wind.

The Extreme Ultraviolet Imager (EUI; \cite{rochus2020}) on board the Solar Orbiter \cite{muller2020}, with its unprecedented high spatial resolution and cadence, provides a good opportunity to probe these unexplored high-frequency waves. Here, we analyze high-cadence EUI observations of polar coronal plumes—open-field, ray-like magnetic structures that act as natural waveguides \cite{Poletto2015}—to search for signatures of previously undetected high-frequency propagating kink waves. Our findings reveal a previously hidden, energetically significant population of high-frequency waves, offering new insights for the acceleration of fast solar wind. While earlier studies occasionally noticed the existence of several high-frequency kink wave events \cite{morton2019,weberg2020,Baweja2025}, this work presents the first large-sample statistical detection of a distinct high-frequency population, provides a systematic assessment of their energy contribution, and places the findings in direct comparison with results from previous instruments.

\section{RESULTS}\label{sec:res}

\begin{figure*}[h]
\centering
\includegraphics[width=1\linewidth]{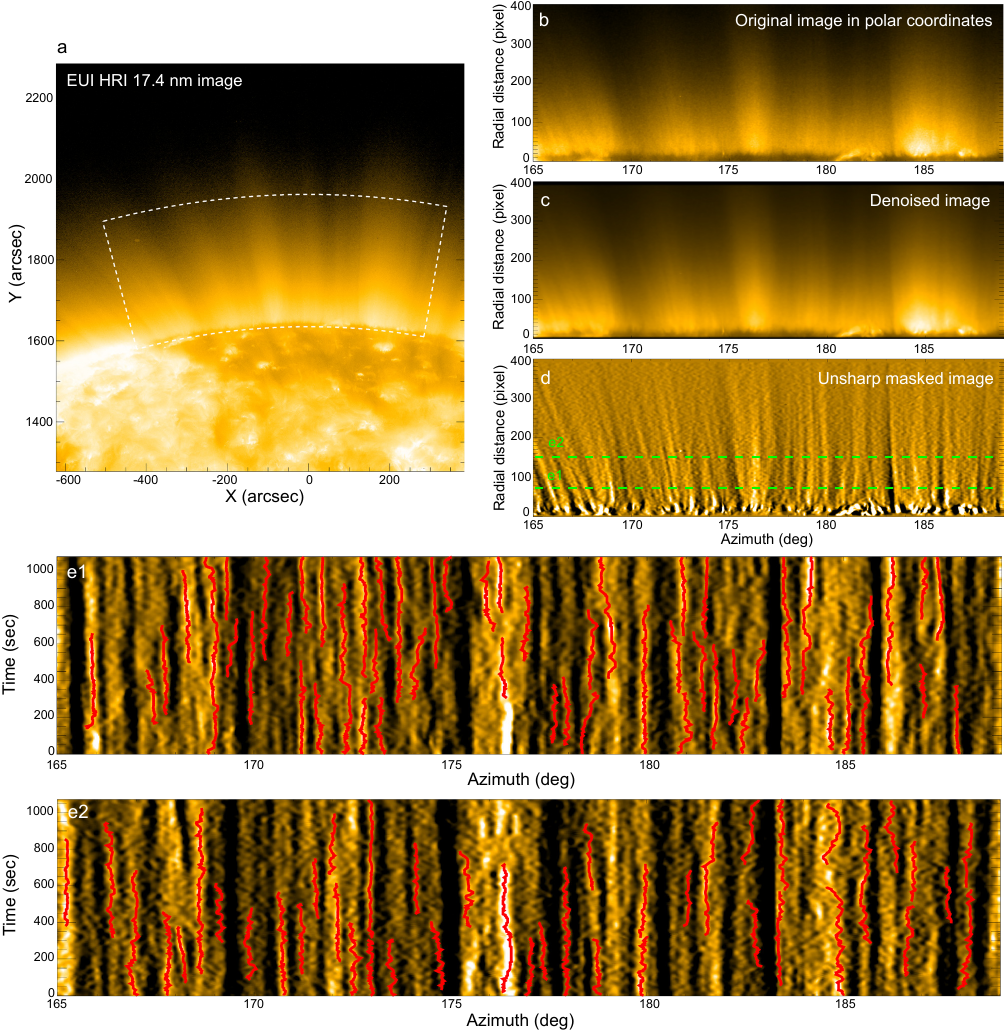}
\caption{Observations by Solar Orbiter/EUI/HRI$_\mathrm{EUV}$ on 2021 September 14. (A) Full field of view showing the north polar coronal hole. The white box outlines the region of interest selected for detailed analysis. (B) The same region after transformation to polar coordinates. (C) Denoised image obtained using the noise-gating technique. (D) Unsharp-masked image highlighting fine plume structures. (E) Two examples of time-distance maps generated along horizontal slits (each with a width of 11 pixels) marked by green dashed lines in panel (D). The blue dotted lines indicate kink wave events identified by the NUWT code, appearing as transversely oscillating threads.}\label{fig:polar}
\end{figure*}

Our analysis uses high–spatial- and temporal-resolution extreme-ultraviolet (EUV) observations at 17.4 nm from the High Resolution Imager (HRI$_\mathrm{EUV}$) of Solar Orbiter/EUI. We focus on the dataset targeting the north polar coronal hole on 2021 September 14 (Figure \ref{fig:polar}(A)). We first aligned the image sequences with cross-correlation method. Then the off-limb region containing plume structures was processed and enhanced to reveal fine-scale threads (Figure \ref{fig:polar}(B–D); see also the `DATA AND METHODS' Section for details). Time–distance (TD) maps were then constructed along 17 horizontal slits at heights from 8 to 41 Mm. These TD maps show prevalent transverse oscillations in plume threads (Figure \ref{fig:polar}(E1) and (E2)), which were automatically identified using the Northumbria University Wave Tracking (NUWT) code \cite{weberg2018}. NUWT also provides the wave period, displacement amplitude, and velocity amplitude via the Fourier transform.

\begin{figure*}[h]
\centering
\includegraphics[width=1\linewidth]{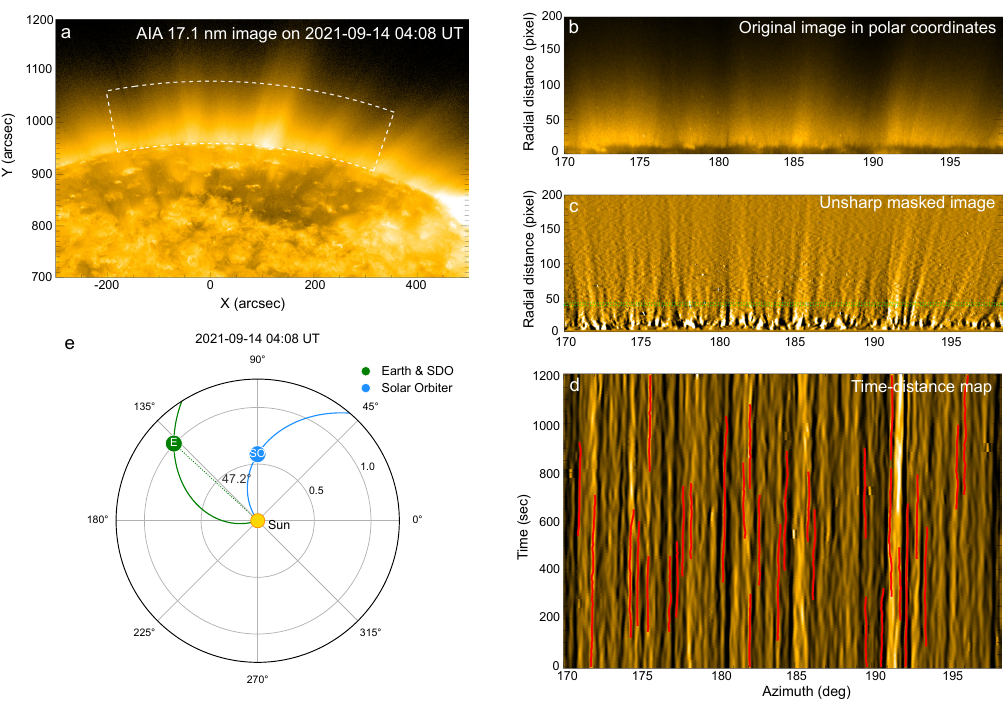}
\caption{Observations by SDO/AIA. (A) Original AIA 17.1 nm image, with the white box marks the region of interest (ROI). (B) The ROI transformed to polar coordinates. (C) Image after denoising and unsharp masking, revealing fine plume structures. The green dotted line indicate an example of slits (with a width of 5 pixels) for TD-map generation. (D) Example TD map showing kink waves detected by NUWT. (E) Heliographic configuration of Solar Orbiter and SDO (close to the Earth) on 2021 September 14.}\label{fig:aia}
\end{figure*}

For comparison, we conducted the same analysis for co-temporal observations from the Atmospheric Imaging Assembly (AIA; \cite{lemen2012}) at 17.1 nm on board the Solar Dynamics Observatory (SDO; \cite{pesnell2012}), as shown in Figure \ref{fig:aia}. This enables a statistical comparison of wave properties across different spatial resolutions and cadences. Key instrumental parameters are summarized in Table \ref{tab:1}. Notably, EUI provides images with spatial resolution and cadence approximately twice as high as those of AIA.

\begin{table}[t]
\begin{tabular}{lll}
\hline
\hline
\textbf{Instruments}             & \textbf{EUI}     & \textbf{AIA}     \\
\hline
Passband                & 17.4 nm & 17.1 nm \\
Cadence (s)             & 5       & 12      \\
Pixel size (Mm)         & 0.21    & 0.44    \\
Resolution (Mm)         & 0.42    & 1.09    \\
\# of Events             & 2318    & 560     \\
Period (s)              & 184±149 & 240±136 \\
Disp. Amp. (km)          & 318±188 & 322±152 \\
Vel. Amp. (\kms)        & 15.4±9.7 & 9.9±5.5 \\
$v_\mathrm{rms}$ (\kms) & 18.2    & 11.3     \\
Energy flux (W m$^{-2}$)      & 30-50  & 11-19  \\
\hline
\end{tabular}
\caption{Instrumental parameters and statistical properties of propagating kink waves detected by Solar Orbiter/EUI and SDO/AIA. Values for period, displacement amplitude, and velocity amplitude are given as the log-normal mean $\pm$ standard deviation. The energy flux ranges are calculated using Equation (\ref{eq:1}) under the stated assumptions.}
\label{tab:1}
\end{table}

\subsection{Statistics of wave parameters}\label{subsec:stat}

\begin{figure*}[ht]
\centering
\includegraphics[width=1\linewidth]{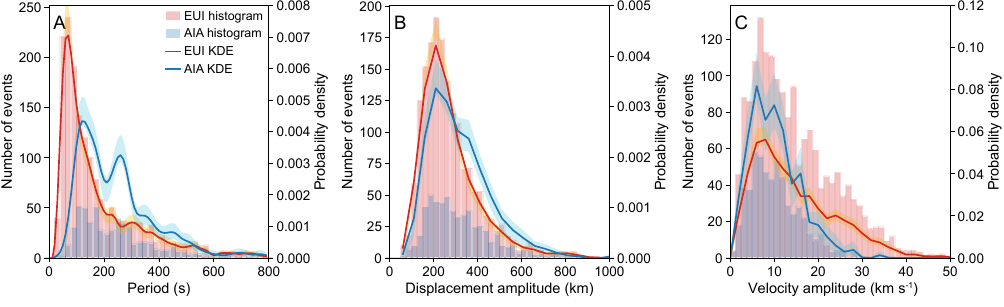}
\caption{Statistical distributions of wave periods, displacement amplitudes, and velocity amplitudes measured from EUI (red) and AIA (blue) observations. Panels A, B, and C show the period, displacement amplitude, and velocity amplitude, respectively. In each panel, the left $y$-axis represents the number of events in each histogram bin, while the right $y$-axis shows the probability density (solid curves) obtained from the kernel density estimate (KDE) with 95\% confidence intervals shaded. Histograms and KDEs share the same $x$-axis for direct comparison.}\label{fig:histogram}
\end{figure*}

Across the 17 EUI TD maps, NUWT detected 2318 wave events (typically 90–130 per slit). In contrast, the AIA observations yielded 560 events over the same time interval. Figure~\ref{fig:histogram} summarizes the statistical distributions of wave period, displacement amplitude, and velocity amplitude for both instruments. To enable a direct comparison despite different sample sizes, we computed probability density estimates using kernel density estimation (KDE) with \texttt{sklearn.neighbors.KernelDensity} from the Python package \texttt{scikit-learn v1.5}. The 95\% confidence intervals (shaded regions) were computed via bootstrap resampling to quantify the uncertainty.

From Figure~\ref{fig:histogram}(A), the period distributions of the two instruments differ substantially, especially at short periods. EUI detects a large population of short-period events, producing a dominant peak at the short periods. A similar feature was also reported by \cite{Baweja2025}, but based on a much smaller sample. Here, 57\% of the EUI events have periods shorter than 150~s, and 38\% are shorter than 100~s, whereas the corresponding fractions for AIA are only 31\% and 9\%. AIA additionally shows a secondary peak near 300 s, consistent with earlier reports of a preferred period of approximately 5 min, likely reflecting photospheric p-mode leakage \cite{tomczyk2007,weberg2020,Morton2025origin}.

The displacement and velocity amplitudes both follow approximately log-normal distributions (Figure \ref{fig:histogram}(B) and (C)), in agreement with previous findings \cite{morton2015,weberg2020}. The displacement amplitudes are broadly similar between the two instruments, with log-normal means of $318 \pm 188$ km (EUI) and $322 \pm 152$ km (AIA). However, the velocity amplitudes derived from EUI exhibit a higher mean value of $15.4 \pm 9.7$ km s$^{-1}$, compared with $9.9 \pm 5.5$ km s$^{-1}$ for AIA. The corresponding root-mean-square (RMS) velocity amplitudes are $v_{\mathrm{rms}}^\mathrm{EUI} = 18.2$ km s$^{-1}$ and $v_{\mathrm{rms}}^\mathrm{AIA} = 11.3$ km s$^{-1}$ (see Table~\ref{tab:1}).

The RMS velocity amplitudes allow a rough estimate of the energy flux carried by the waves \cite{tvd2014,morton2015,weberg2018,morton2019,Morton2025NA}:
\begin{equation}\label{eq:1}
F = f \langle\rho\rangle v_\mathrm{rms}^2 c_\mathrm{k}\,,
\end{equation}
where $f$ is the filling factor, $\langle\rho\rangle$ is the mean density, and $c_\mathrm{k}$ is the propagation speed of kink waves. Following previous works \cite{McIntosh2011,morton2015,weberg2018,morton2019,yang2020Sci,yang2024,Morton2025NA}, the EUI and AIA wave energy fluxes are estimated to be 30–50 W m$^{-2}$ and 11–19 W m$^{-2}$, respectively. These values are subject to large uncertainties associated with the adopted physical parameters and assumptions (see the `Uncertainties in the energy flux estimation' Section for details). Importantly, irrespective of absolute scale, the EUI-derived energy flux is larger than AIA’s by a factor of approximately 2.6, highlighting the enhanced energy contribution of high-frequency waves resolved only by EUI’s higher cadence and spatial resolution.

\subsection{Power spectrum}\label{subsec:psd}

\begin{figure}[ht]
\centering
\includegraphics[width=1\linewidth]{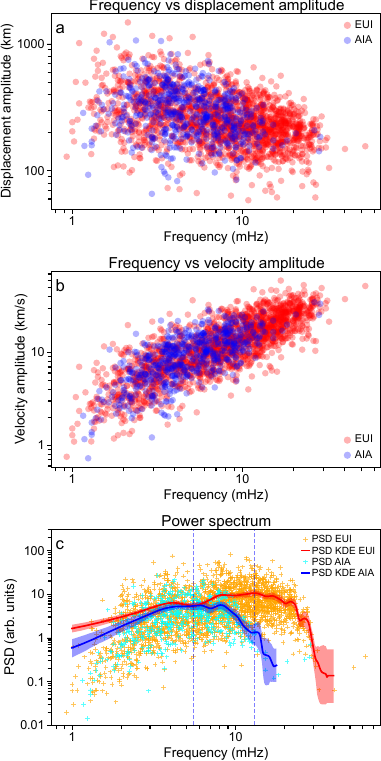}
\caption{Frequency distribution of wave properties. Panel A shows the scatter plot between frequency and displacement amplitudes. Panel B shows the scatter plot between frequency and velocity amplitudes. Panel C shows the power spectral density (PSD) for individual wave events. Solid curves show the mean PSD obtained by kernel regression at each frequency, with shaded regions indicating the 95\% confidence intervals (CIs). The vertical dashed lines correspond to 5.5 mHz and 13.0 mHz.}
\label{fig:psd}
\end{figure}

In Figure \ref{fig:psd}, we further considered the frequency distribution of wave properties. 
Figure \ref{fig:psd}(a) and (b) present the scatter plots of amplitude versus frequency.
The displacement amplitude decreases slightly with increasing frequency, consistent with \cite{morton2019} and \cite{weberg2020}, while the velocity amplitude increases with frequency. The trends shown here are consistent for both instruments, while the EUI results appear to naturally extend these trends to higher frequencies. This demonstrates that high-frequency waves (typically with relatively small displacement amplitudes and large velocity amplitudes) are more effectively resolved in the high-cadence, high-resolution EUI data, which further contributes to the higher energy fluxes inferred from EUI.

The power spectrum of detected waves can help us quantify energy distribution with frequency and provide insights about the wave origin \cite{tomczyk2007,Lim2024,Morton2025origin}. 
Figure~\ref{fig:psd}(c) presents the power spectra derived from the EUI and AIA observations. The analysis follows the method described in \cite{morton2019}. For each detected wave event, the power spectral density (PSD) is calculated as the product of the mean square velocity amplitude and the normalized occurrence probability at its frequency. The latter is derived from the KDE described above. The resulting PSDs for individual events are shown as scatter points in Figure~\ref{fig:psd}(c). We note that although individual high-frequency events tend to have larger velocity amplitudes (Figure~\ref{fig:psd}(b)), their lower occurrence probability causes the overall PSD to decline toward the highest frequencies.

To obtain a smooth and unbiased representation of the PSD, we fitted the result using a Nadaraya–Watson kernel regression. The fitting results are shown as red and blue solid curves. The shaded regions represent the 95\% confidence intervals (CIs), estimated from 1000 bootstrap resamples of the input data. Additional methodological details are provided in the Supplementary Material.

The power spectra reveal that the EUI and AIA curves are broadly comparable at low frequencies below around 7 mHz. At higher frequencies, however, EUI exhibits substantially more power, especially above 10 mHz (periods shorter than 100 s). 
These characteristics indicate that EUI can effectively detect abundant high-frequency waves carrying substantial energy, which are largely unresolved in AIA observations due to its lower spatial and temporal resolution. A resolution-degradation test further demonstrating the impact of the spatial and temporal resolution can be found in the Supplementary Materials.

To further quantify the energy distribution, we integrated the spectra over two representative frequency bands. For EUI, the integrated power in the high-frequency range (10–30 mHz) exceeds that in the low-frequency range (2–10 mHz) by more than a factor of two. This verifies that the enhanced high-frequency power detected by EUI constitutes a substantial fraction of the total wave energy in coronal plumes.

\section{Discussion}

\subsection{Potential origins of high-frequency waves}

The key finding of this study is the detection of an abundant population of high-frequency kink waves carrying a substantial fraction of the coronal wave energy flux. While a detailed observational investigation of their generation mechanisms is beyond the scope of this work, it is worthwhile to discuss several plausible origins.

The first possibility is interchange reconnection between closed and open magnetic field lines, widely considered a source of transverse waves (e.g., \cite{tian2014,YangLP2025}). Theoretical and numerical studies suggest that magnetic field reconfiguration following reconnection naturally generates transverse waves, with wavelengths set by the spatial scales of the reconnecting structures, particularly the lengths of closed loops \cite{Lynch2014,Morton2025origin}. Assuming that the wavelength roughly matches twice the loop length $L$, the wave period can be estimated as $P \sim 2L/c$. Observations of magnetic loops in quiet-Sun regions and coronal holes typically yield $L \sim 5$–40~Mm \cite{gao2022,Madjarska2024,Shrivastav2024,Lim2024}. With a propagation speed $c = 300$–500~km~s$^{-1}$ \cite{yang2020,yang2020Sci,yang2024}, this corresponds to periods of 20–266~s (frequencies of 3.8–50~mHz), consistent with the high-frequency range observed in the EUI data.

The second possibility is that transverse waves in plumes are driven by spicules at their bases. Kink waves with periods below 100~s are frequently reported in spicules \cite{He2009, bate2022}. Meanwhile, spicules are often observed to be closely associated with the plumes above them \cite{Jiao2015,Samanta2015}. Recently, Qi et al. \cite{Qi2026} analyzed EUI observations of polar plumes and identified a direct connection between kink waves in a plume and kink waves in a spicule at its base exhibiting similar periods. Their numerical simulations further demonstrate that lateral motions of spicules can naturally excite kink waves in the overlying plumes sharing the same periods. This scenario suggests that at least some of the high-frequency waves detected here could be driven by spicule motions at their bases.

The third possibility is turbulent cascading. Reflection of low-frequency transverse waves, or the interaction of counter-propagating waves, can generate a turbulent cascading to higher frequencies \cite{tvd2020SSRv}.
Observationally, both outward and inward propagating kink waves have been reported in the open-field corona \cite{Liu2014,morton2015}. In this work, we noticed a slight decrease of wave periods with height (see Figure S1 in the Supplementary Material), which might also indicate partial wave reflection (see also \cite{weberg2020}). Even for non-reflecting waves, transverse structuring can induce self-cascade through a nonlinear process known as uniturbulence \cite{magyar2017,tvd2020ApJ}. These mechanisms may facilitate turbulent cascading from lower-frequency waves to frequencies exceeding 10~mHz. 

There are also other possible origins. For example, high-frequency wave power may be generated in the photosphere and tunnel upward into the upper atmosphere \cite{Skirvin2024}, or some power may leak from high-harmonic chromospheric resonant oscillations \cite{Jess2020}. Moreover, recent simulations have shown that nonlinear longitudinal-to-transverse mode conversion near the equipartition layer can produce transverse waves with frequencies of $\sim$20 mHz \cite{shoda2018, kuniyoshi2024}. Distinguishing between these possibilities will require future studies combining high-resolution imaging, spectroscopy, and magnetic field measurements to trace waves from their low atmospheric origin into the corona.

\subsection{Comparisons with previous findings}

Very recently, Baweja et al. \cite{Baweja2025} also analyzed transverse waves in polar coronal plumes using Solar Orbiter/EUI data, focusing on the coexistence of slow-mode waves and kink waves. Using manual identification, they reported 98 transverse wave events extending up to approximately 20 Mm. Their measured wave parameters and height-dependent trends are broadly consistent with the present study. However, our work provides a significantly larger statistical sample (2318 events versus 98) thanks to the automated NUWT algorithm, enabling more robust statistical analysis and power spectral characterization. Moreover, our detections extend to larger heights (up to 41 Mm). Finally, we present a direct statistical comparison with co-temporal SDO/AIA observations of the same coronal hole, which strongly supports that the superior cadence and spatial resolution of EUI allow the detection of a previously hidden population of high-frequency waves and their associated energy flux.

The power spectrum of propagating kink waves has been investigated in several previous studies. Our EUI observations reveal some distinct characteristics compared with earlier results. Spectra obtained from Doppler velocity measured by the Coronal Multi-Channel Polarimeter (CoMP; \cite{tomczyk2008}) and the Cryogenic Near-Infrared Spectropolarimeter (Cryo-NIRSP; \cite{Fehlmann2023}) on the Daniel K. Inouye Solar Telescope (DKIST; \cite{DKIST2020})  generally exhibit power laws with negative indices between –2.0 and –0.5, accompanied by a bump around 3.5–4 mHz \cite{tomczyk2007, morton2019, Morton2025NA}. In contrast, although the 3.5-4 mHz bump can still be roughly reproduced by our EUI observations, our spectrum do not follow such a power-law behavior. Instead, our results below 10 mHz are more consistent with previous studies using SDO/AIA data \cite{morton2019}. The discrepancy between our results and those obtained with CoMP and DKIST/Cryo-NIRSP may be attributed to differences between spectroscopic and imaging observations. For example, the Doppler velocity measurements from spectroscopy generally leads to an underestimate of amplitude due to line-of-sight averaging effect \cite{pant2019,shi2021forward,Morton2025NA}.

The main difference between EUI and AIA results occurs at high frequencies ($\ge$7 mHz), which can be explained by the different spatial resolutions and cadences. Noticeably, recent DKIST/Cryo-NIRSP observations \cite{Morton2025NA} can also generate time-distance map of intensity along the spectrograph slit, analogous to our Figure \ref{fig:polar}(E). This allows similar wave analysis to be applied to DKIST/Cryo-NIRSP data. One might therefore expect a similar power spectrum from these data, given Cryo-NIRSP's higher resolution (0.6") and cadence (1 s) than our EUI dataset.  However, \cite{Morton2025NA} reported a power spectrum that differs from ours and instead resembles their Doppler-velocity–based spectroscopic measurements—specifically, showing no strong power in the 10–20 mHz range, but a significant decrease in power (see their Figure 5c). Earlier CoMP and AIA studies \cite{tomczyk2007, morton2019} likewise show such power decreases over 10 mHz, in contrast to our EUI findings presented in Figure \ref{fig:psd}(C). 

Several physical and observational factors may explain this discrepancy:
(1) The Cryo-NIRSP observations analyzed by \cite{Morton2025NA} were taken at a height of $\sim$70 Mm above the solar surface using the Fe \textsc{xiii} 1074.7 nm line, which forms at $\sim$1.6 MK. This region is both higher in the corona and hotter than that observed in our study (which uses the 17.4 nm EUV band, sensitive to $\sim$1 MK plasma, to observe the corona below 42 Mm). High-frequency waves may be more efficiently damped (e.g., via resonant absorption) before reaching the hotter, higher layer observed by Cryo-NIRSP. (2) The Cryo-NIRSP data were obtained in coronal open-field regions adjacent to low-latitude active regions, whereas our analysis focuses on polar coronal holes. Differences in the underlying atmospheric environment may also influence the observed wave properties. (3) Subtle differences in wave-detection algorithms, significance thresholds, or background subtraction between studies could also influence the resulting power spectrum. Therefore, we do not regard the differences between the EUI and Cryo-NIRSP spectra as undermining either result. Instead, they likely reflect the dependence of wave power on coronal height, plasma temperature, and magnetic environment. Future coordinated observations at multiple heights and across diverse magnetic environments, combined with cross-instrument analyses, will be essential to deepen our understanding of high-frequency waves in the corona.

There are some other interesting features worth discussing in our EUI power spectrum. As indicated by the dashed lines in Figure \ref{fig:psd}(C), the spectrum exhibits a local minimum at $ \sim $5.5 mHz and a turning point near 13 mHz. The local minimum at $ \sim $5.5 mHz has also been identified in DKIST/Cryo-NIRSP observations \cite{Morton2025NA}. Following their interpretation, the power spectrum shows distinct peaks corresponding to two components of different origins: a narrow peak near 4 mHz from mode conversion of the photospheric p-modes (see also \cite{Khomenko2012,Miriyala2025}) and an additional peak near 6 mHz attributed to uncertain origin processes (possibly related to magnetic reconnection or chromospheric 3-min oscillations). Their combination could naturally produce a local minimum between 4 mHz and 6 mHz. Meanwhile, the AIA spectrum does not show this minimum, likely owing to its lower spatial–temporal resolution and smaller sample size. 

The apparent turning point near 13~mHz should be interpreted with caution. One possible physical interpretation is that this frequency range reflects relatively efficient wave excitation, for example through interchange reconnection. For typical loop lengths of 10--20~Mm in coronal bright points \cite{Madjarska2024,Lim2024} and kink speeds of 300--500~km~s$^{-1}$, the corresponding periods are broadly comparable to the observed range. However, the current data do not allow us to determine whether this apparent feature represents a genuine preferred injection frequency. An alternative explanation is that the apparent turning point is partly produced by observational and methodological limitations. The measured power spectrum is affected by finite cadence, finite time-series length, wave-detection efficiency, and measurement noise. In practice, the intrinsic wave spectrum is convolved with the observational response and noise properties, so apparent peaks, dips, or changes in slope may arise even when the underlying spectrum is smooth. Consistent with this possibility, when the EUI data are degraded to AIA's resolution, the apparent turning point shifts downward to $\sim$10~mHz (see Figure S2 in the Supplementary Material). This suggests that the location of such features may depend on the effective temporal and spatial resolution of the observations. Future analysis of multiple EUI datasets, together with forward modeling of the observational selection effects, will be required to test whether these spectral features are physical or specific to the current dataset.

\subsection{Uncertainties in the energy flux estimation}

When applying Equation (\ref{eq:1}) to estimate the energy flux, we adopted $f=0.5$, $\langle\rho\rangle = 3\times10^{-13}\,\mathrm{kg~m^{-3}}$, and $c_\mathrm{k}=300$–$500$ \kms. Meanwhile, the velocity amplitudes are taken from Table \ref{tab:1}, and are further multiplied by $\sqrt{2}$ to account for the random polarization of propagating kink waves \cite{thurgood2014,weberg2018,Morton2025NA}, since measured velocities only represent the plane-of-sky component. The choice of  $ f $, $ \langle\rho\rangle $, and $ c_\mathrm{k} $, together with their associated uncertainties, is discussed below.

The filling factor $ f $
is difficult to measure directly due to projection effects and line-of-sight superposition in EUV imaging observations. Previous studies have used values ranging from $ f=1 $ (e.g., \cite{McIntosh2011,Morton2025cycle}) to $ f\sim0.5 $ (e.g., \cite{morton2015,morton2019}). In this work, we adopt $ f=0.5 $, a widely used and conservative choice for imaging-based studies, to allow consistent comparison with earlier results. However, Equation (\ref{eq:1}) is strictly valid only for small filling factors (
$ f\lesssim10\% $; \cite{tvd2014}). Stereoscopic observations by Huang et al. \cite{Huang2021} further suggest a lower limit of 
$ f\approx2–3.4\% $ for coronal plumes. If the actual filling factor lies in the range 5–10\%, the absolute energy flux would be reduced by a factor of 5–10 relative to the values presented in Table \ref{tab:1}. Importantly, this correction applies equally to all previous estimates using the same formulation, so the relative enhancement of energy flux measured by EUI remains robust. We also note that plumes may expand with height into funnel-like structures that cover a large fraction of the coronal-hole area \cite{Tu2005,Moore2023,Alzate2025}, suggesting the effective filling factor could increase with height. A precise determination of $ f $ requires further investigation.

For the kink speed $ c_\mathrm{k} $, the range 
300–500~km~s$ ^{-1} $ is consistent with CoMP measurements  \cite{morton2015,yang2020Sci,yang2020,yang2024} and the value ($ 420\pm 80 $~\kms) used in \cite{morton2019}. However, previous CoMP-based spectroscopic measurements typically start above 1.05-1.10 solar radii ($ R_\odot $) and cannot constrain kink speeds at low coronal heights ($ \lesssim $42 Mm) studied here. We therefore adopt this relatively broad range for robustness. For the mean density $ \langle\rho\rangle $, we adopt a conservative value of $ 3\times10^{-13}\,\mathrm{kg~m^{-3}} $, consistent with multiple estimates near 1.01–1.06$ R_\odot $ (7–42~Mm)\cite{morton2015,Long2023}.

The potential impact of nonlinear effects on energy flux estimation is also considered here. Some wave events with large velocity amplitudes may enter the nonlinear regime, and our flux estimates are based on linear MHD theory.  Statistically, however, nonlinear effects are likely subdominant compared with uncertainties in density and propagation speed. Given the typical kink speed of 
300–500~km~s$ ^{-1} $, only 8.6\% of events have velocity amplitudes above 30~km~s$ ^{-1}$  and just 
0.3\% exceed 
50~km~s$ ^{-1} $. Therefore, large-amplitude events (exceeding 10\% of the wave phase speed) are quite rare to alter the overall flux significantly. Nevertheless, nonlinear effects, such as uniturbulence, may still develop over sufficient propagation distances even for small amplitudes \cite{tvd2020ApJ}. To our knowledge, no direct study has quantified nonlinear corrections to energy flux for propagating kink waves. For standing kink waves, Guo et al. \cite{guo2023flux} used numerical simulations to derive a nonlinear correction factor of about 2 relative to the linear formula, meaning that linear estimates tend to underestimate the true flux. For propagating waves, the exact correction remains unconstrained, but it is plausible that Equation (\ref{eq:1}) also yields a conservative, lower-bound estimate.

Additional uncertainty arises from the temporal intermittency of wave activity: individual wave events persist for only a fraction of the total observing time ($T_\mathrm{tot}$). Accounting for temporal intermittency using an effective RMS velocity $v_\mathrm{rms, eff}^2 = \langle v_i^2 T_i / T_\mathrm{tot} \rangle$  ($v_i$ and $T_i$ stands for the velocity amplitude and duration for the $i$-th event) reduces both estimates by roughly 30\%–40\%, though this correction is itself uncertain due to possible overlap of threads in the TD maps.

Considering all these factors, the resulting flux may still be insufficient to explain coronal heating and fast solar wind acceleration, which likely requires the energy flux of around 500 W m$^{-2}$ according to recent estimates \cite{HuangZG2024,Morton2025cycle}. This implies that additional contributions from even higher-frequency waves (up to $ \sim $1 Hz; see \cite{YangLP2025}) and torsional Alfv\'{e}n waves \cite{Morton2025NA} are likely needed. Future high-resolution observations from a polar vantage point and dedicated measurements of wave speed and density will help reduce these uncertainties.

\section{CONCLUSION}

In this work, we analyzed propagating kink waves in coronal plumes above a polar coronal hole using Solar Orbiter/EUI and SDO/AIA observations. Our focus was on the previously hidden population of high-frequency waves, which may not have been detected in observations with lower cadences and resolutions. We identified 2318 wave events in EUI data and 560 events in AIA data. Statistical analysis shows that a significant fraction of these waves observed by EUI have high frequencies ($\ge$10 mHz). The power spectrum derived from EUI also reveals that high-frequency waves carry substantial energy flux, which is largely undetectable in AIA observations. Possible origins of these high-frequency waves include interchange reconnection, spicule oscillations at the plume base, and turbulent cascading.

Similarly, EUI observations have enabled the detection of high-frequency kink oscillations in small coronal loops \cite{Lim2024,Shrivastav2024,Shrivastav2025}. Although those waves are primarily interpreted as standing modes, different from the propagating modes studied here, the occurrence of high-frequency transverse motions in diverse coronal structures further highlights the importance of high-resolution, high-cadence observations.

Theoretical and numerical studies indicate that high-frequency propagating kink waves can be more efficiently damped and dissipated than low-frequency waves through several processes. Resonant absorption is a well-established damping mechanism for propagating kink waves, in which wave energy is transferred to the resonant layer within the inhomogeneous boundary between the inside and outside of the flux tube, generating azimuthal Alfv\'{e}n waves \cite{terradas2010, Goossens2011, Gao2024pkw}. The damping length due to resonant absorption scales inversely with frequency (proportional to $1/f$; e.g., \cite{terradas2010}), so higher-frequency waves are expected to have shorter damping lengths. Although resonant absorption does not directly dissipate wave energy, it can enhance phase mixing in the boundary layer, cascading energy to small scales and ultimately producing heating via Ohmic and/or viscous dissipation \cite{Pagano2019, Gao2024pkw}. Therefore, high-frequency kink waves are likely to be particularly relevant for energy deposition in the lower corona.

Another potentially important nonlinear mechanism is uniturbulence, which can occur when kink waves propagate through transversely structured plasma \cite{magyar2017}. Unlike resonant absorption, the associated cascade rate does not follow a simple $1/f$ dependence, but depends on parameters such as wave amplitude, transverse density contrast, and the characteristic width of the waveguide \cite{tvd2020ApJ}. Although this mechanism is not intrinsically favored by higher frequency, the high-frequency waves detected here tend to have relatively large velocity amplitudes (Figure 4(B)), which could make nonlinear self-cascade more efficient. Uniturbulence may therefore provide an additional pathway for converting the observed transverse wave energy into smaller scales and eventually heat, especially in the highly structured plume environment of the lower corona.

At larger heliocentric distances, ion cyclotron resonance may provide a further channel for dissipating Alfv\'{e}nic fluctuations \cite{Marsch1997}, although the waves detected here are still far below the ion cyclotron frequency in the low corona. Thus, we do not claim that the observed 10--50 mHz waves can be directly dissipated by cyclotron resonance in the lower corona. Instead, their significance is that they substantially reduce the frequency gap that must be bridged by turbulent cascade before wave power can reach ion-cyclotron scales. Earlier studies suggested that transverse waves with periods of order 50 s could undergo cyclotron-resonant damping at large heliocentric distances, around $60R_\odot$ \cite{He2009}. The abundant high-frequency wave population detected in this work therefore provides observational support for the existence of wave power at frequencies closer to the range required by such extended-corona and inner-heliospheric dissipation scenarios. 

These mechanisms may operate complementarily across different height ranges. In the lower corona, resonant absorption, phase mixing, and nonlinear self-cascade through uniturbulence may redistribute or dissipate part of the high-frequency wave energy locally. The remaining wave power may propagate outward, where turbulent cascade could transfer energy to progressively smaller temporal and spatial scales and eventually enable kinetic dissipation processes such as cyclotron resonance or Landau-type wave--particle interactions. Because fast solar wind acceleration occurs over a broad range of heliocentric distances, a combination of local coronal damping and extended-heliospheric turbulent/kinetic dissipation is likely required. Future coordinated remote-sensing and in-situ observations, together with forward modelling and numerical simulations, will be essential for determining the relative importance of these mechanisms.


Future work could also extend the wave analysis to additional EUI datasets of polar coronal holes to study PKW properties across multiple regions. Looking ahead, the forthcoming SPO mission will provide continuum and EUV imaging, magnetograms, and in-situ solar wind measurements from a polar vantage point for the first time. Such polar observations will be subject to less LOS superposition effect compared to AIA and EUI observations. Thus, SPO observations will likely enable direct identification of PKW origins in polar coronal holes and further clarify their role in the acceleration of fast solar winds.

\section{DATA AND METHODS}\label{sec:method}

The primary dataset of this research is from the Solar Orbiter/EUI/HRI$_\mathrm{EUV}$, which is generally referred as EUI throughout the text. EUI acquired 225 images in the 17.4 nm passband between 04:08 and 04:26 UT on 2021 September 14, when the Solar Orbiter was at a heliocentric distance of 0.59 au. This dataset has a cadence of 5 s, an exposure time of 2.8 s, and a pixel size of 0.492" ($\sim$0.21 Mm). Figure \ref{fig:polar}(A) displays the full field of view of EUI/HRI$_\mathrm{EUV}$. Our analysis focused on the off-limb region above the coronal hole, where plumes and their transverse motions are most clearly visible. The image sequence was first aligned via cross-correlation using stable on-disk sub-regions to remove instrumental jitter \cite{Yuan2018,Lim2024}. For this alignment, we selected relatively stable sub-regions on the solar disk (the lower part in Figure \ref{fig:polar}(A)). The aligned images were then transformed into polar coordinates, as shown in Figure \ref{fig:polar}(B) (corresponding to the white box in panel A). Because the coronal hole emission is faint and suffers from a low signal-to-noise ratio (SNR), we applied the noise-gating technique developed by DeForest \cite{DeForest2017} to suppress background noise. The denoised image is shown in Figure \ref{fig:polar}(C).

To further enhance fine-scale plume structures, we performed unsharp masking, which was widely used in previous studies \cite{McIntosh2011,morton2015,weberg2018,weberg2020}. Specifically, we produced a Gaussian-smoothed version of each denoised image using a kernel size of 41 pixels ($\sim$8.6 Mm), larger than typical plume substructure widths (e.g., \cite{DeForest2018}). Subtracting the smoothed image from the denoised one yielded the unsharp-masked image (Figure \ref{fig:polar}(D)), which clearly reveals multiple fine structures. This process was applied to all 225 images to produce a three-dimensional data cube for generating TD maps, where transverse motions appear as fluctuating intensity tracks.

TD maps were constructed along horizontal slits spaced every 10 pixels ($\sim$2.1 Mm) from 7.6 to 41.2 Mm above the solar surface. Each slit has a width of 11 pixels ($\sim$2.3 Mm) to improve SNR. In total, 17 TD maps were generated, with two of which are shown in Figure \ref{fig:polar}(E1) and (E2). In these TD maps, numerous plume threads display transverse oscillations. These wave events were automatically identified and analyzed using the NUWT algorithm. In brief, NUWT detects local intensity maxima associated with overdensed plume structures using Gaussian fitting, and links them in time to form continuous displacement time series (``threads"). A fast Fourier transform is then applied to each thread, and significant wave components are selected using a white-noise–based statistical significance test, from which the periods,  displacement amplitudes and velocity amplitudes are derived. For a more detailed description, please refer to \cite{weberg2018}. 

Similar to previous observations \cite{weberg2018,weberg2020,Morton2025NA}, the detected threads often show a multi-frequency nature. During the NUWT analysis, the Fourier transform is applided for each oscillating thread and multiple significant periodicities or frequencies can be identified in a single event, based on the criteria that the Fourier power is larger than 95\% significance level. Some examples of threads and corresponding Fourier spectra are shown in the Supplementary Material.

We note that automated detection algorithms inevitably have some limitations. For example, low SNR can introduce spurious intensity maxima, potentially leading to problematic thread identification. To mitigate these effects as much as possible, we first pre-processed the images to enhance the SNR (e.g., by constructing TD maps using 5-pixel-wide slits and applying the noise-gating technique to suppress noise). In addition, we imposed a strict criterion for thread identification, requiring a minimum thread length of 60 time steps (300 s). While we cannot completely rule out the possibility that a small fraction of the detected oscillation events may be less robust, the overall statistical results should remain reliable. Furthermore, similar high-frequency transverse waves have been independently detected via manual identification in \cite{Baweja2025}, providing additional support for our findings.

We also analyzed SDO/AIA data at 17.1 nm channel which  samples plasma at a temperature slightly lower than that of the EUI 17.4 nm band \cite{Chen2021}. This channel have been widely used in previous studies of plume waves \cite{McIntosh2011,morton2015,morton2019,weberg2020}. A comparison between AIA and EUI is thus instructive, as EUI provides significantly higher spatial and temporal resolution (0.21 Mm, 5 s) than AIA (0.44 Mm, 12 s). The instrument information of AIA and EUI is also compared in Table \ref{tab:1}.

We applied a similar analysis pipeline to AIA data targeting the same polar coronal hole (see Figure \ref{fig:aia}) for the similar time interval (04:07-04:27 UT). Parameters were adjusted to ensure consistency between the two instruments. For instance, the slit width for AIA was set to 5 pixels ($\sim$2.2 Mm), matching the width used for EUI (11 pixels or $\sim$2.3 Mm). We also note that the Solar Orbiter–SDO separation angle was about 48° (Figure \ref{fig:aia}(E)). The difference in viewing angles means that two instruments may sample different subsets of plumes and project transverse displacements onto different planes of sky. Consequently, direct one-to-one correspondence between detected waves is not valid. Nevertheless, the statistical comparison of wave properties between the two datasets remains scientifically meaningful, as projection effects are expected to average out over a large sample (e.g., \cite{pant2020}).

\section{DATA AVAILABILITY}

The data analysed during the current study are obtained from EUI Data Release 4.0 (https://doi.org/10.24414/s5da-7e78) and from the Joint Science Operations Center (JSOC) database (http://jsoc.stanford.edu/). 

\section{Funding}

This work was supported by NSFC grant 12425301, National Key R\&D Program of China No. 2022YFF0503800, Strategic Priority Research Program of the Chinese Academy of Sciences (Grant No. XDB0560000), China's Space Origins Exploration Program, and Specialized Research Fund for State Key Laboratory of Solar Activity and Space Weather. H.T. also acknowledges support from the New Cornerstone Science Foundation through the XPLORER Prize. Solar Orbiter is a space mission of international collaboration between ESA and NASA, operated by ESA. The EUI instrument was built by CSL, IAS, MPS, MSSL/UCL, PMOD/WRC, ROB, LCF/IO with funding from the Belgian Federal Science Policy Office (BELSPO/PRODEX PEA 4000134088, 4000112292 and 4000106864). TVD was supported by a Senior Research Project (G088021N) of the FWO Vlaanderen. Furthermore, TVD received financial support from the Flemish Government under the long-term structural Methusalem funding program, project SOUL: Stellar evolution in full glory, grant METH/24/012 at KU Leuven. The research that led to these results was subsidised by the Belgian Federal Science Policy Office through the contract B2/223/P1/CLOSE-UP. It is also part of the DynaSun project and has thus received funding under the Horizon Europe programme of the European Union under grant agreement (no. 101131534). Views and opinions expressed are however those of the author(s) only and do not necessarily reflect those of the European Union and therefore the European Union cannot be held responsible for them. DL thanks the Belgian Federal Science Policy Office (BELSPO) for the provision of financial support in the framework of the PRODEX Programme of the European Space Agency (ESA) under contract number 4000143743. MG acknowledges the support from the National Natural Science Foundation of China (12203030), the Taishan Scholars Program Special Fund (tsqn202408051) and the Shandong Provincial Natural Science Foundation for Excellent Young Scientists Program, Overseas (grant No. 2025HWYQ-019). We also acknowledge support by ISSI/ISSI-BJ to the team "Small-scale Eruptions on the Sun".

Conflict of interest statement. None declared.

\section{AUTHOR CONTRIBUTIONS}

H.T. and Y.G. initiated this study. Y.G. analyzed the data, plotted the figures, and wrote the initial manuscript under the guidance of H.T. and T.V.D.. R.M. contributed to the data analysis. D.L. provided instructions in processing the Solar Orbiter/EUI data. All authors  discussed the results and contributed to the writing of the manuscript.

\appendix

\bibliographystyle{nsr}
\bibliography{nsr_sample}

@ARTICLE{McIntosh2011,
       author = {{McIntosh}, Scott W. and {de Pontieu}, Bart and {Carlsson}, Mats and {Hansteen}, Viggo and {Boerner}, Paul and {Goossens}, Marcel},
        title = "{Alfv{\'e}nic waves with sufficient energy to power the quiet solar corona and fast solar wind}",
      journal = {Nature},
         year = 2011,
        month = jul,
       volume = {475},
       number = {7357},
        pages = {477-480},
          doi = {10.1038/nature10235},
       adsurl = {https://ui.adsabs.harvard.edu/abs/2011Natur.475..477M}
}

@ARTICLE{Qi2026,
	author = {{Qi}, Youqian and {Guo}, Mingzhe and {Huang}, Zhenghua and {Van Doorsselaere}, Tom and {Li}, Bo and {Xia}, Lidong and {Wei}, Hengyuan and {Fu}, Hui},
	title = "{Propagating Kink Waves in Chromospheric Jetlike Structures and Coronal Plumelets}",
	journal = {the Astrophysical Journal},
	year = 2026,
	month = apr,
	volume = {1001},
	number = {2},
	eid = {173},
	pages = {173},
	doi = {10.3847/1538-4357/ae5792},
	archivePrefix = {arXiv},
	eprint = {2603.24892},
	primaryClass = {astro-ph.SR},
	adsurl = {https://ui.adsabs.harvard.edu/abs/2026ApJ..1001..173Q}
}

@ARTICLE{Shrivastav2025,
	author = {{Shrivastav}, Arpit Kumar and {Pant}, Vaibhav and {Kumar}, Rohan and {Berghmans}, David and {Van Doorsselaere}, Tom and {Banerjee}, Dipankar and {Petrova}, Elena and {Lim}, Daye},
	title = "{On the Existence of Long-period Decayless Oscillations in Short Active Region Loops}",
	journal = {the Astrophysical Journal},
	year = 2025,
	month = jan,
	volume = {979},
	number = {1},
	eid = {6},
	pages = {6},
	doi = {10.3847/1538-4357/ad95fb},
	archivePrefix = {arXiv},
	eprint = {2411.15646},
	primaryClass = {astro-ph.SR},
	adsurl = {https://ui.adsabs.harvard.edu/abs/2025ApJ...979....6S}
}

@ARTICLE{DeForest2017,
	author = {{DeForest}, C.~E.},
	title = "{Noise-gating to Clean Astrophysical Image Data}",
	journal = {the Astrophysical Journal},
	year = 2017,
	month = apr,
	volume = {838},
	number = {2},
	eid = {155},
	pages = {155},
	doi = {10.3847/1538-4357/aa67f1},
	archivePrefix = {arXiv},
	eprint = {1703.06228},
	primaryClass = {astro-ph.IM},
	adsurl = {https://ui.adsabs.harvard.edu/abs/2017ApJ...838..155D}
}

@Article{Shang2025,
	title = {Statistical study of auroral variability under different solar wind conditions based on classification using deep learning techniques},
	journal = {Earth and Planetary Physics},
	volume = {9},
	number = {6},
	pages = {1163-1170},
	year = {2025},
	issn = {2096-3955},
	doi = {10.26464/epp2025075},	
	url = {https://www.eppcgs.org/en/article/doi/10.26464/epp2025075},
	author = {ZhiYuan Shang and ZhongHua Yao and Jian Liu and LinLi Xu and Yan Xu and BinZheng Zhang and RuiLong Guo and Yuan Yu and Yong Wei}
}

@ARTICLE{Zhao2011,
	author = {{Zhao}, Hong and {Zong}, QiuGang and {Wei}, Yong and {Wang}, YongFu},
	title = "{Influence of solar wind dynamic pressure on geomagnetic Dst index during various magnetic storms}",
	journal = {Science in China E: Technological Sciences},
	year = 2011,
	month = jun,
	volume = {54},
	number = {6},
	pages = {1445-1454},
	doi = {10.1007/s11431-011-4319-y},
	adsurl = {https://ui.adsabs.harvard.edu/abs/2011ScChE..54.1445Z}
}

@ARTICLE{tomczyk2007,
       author = {{Tomczyk}, S. and {McIntosh}, S.~W. and {Keil}, S.~L. and {Judge}, P.~G. and {Schad}, T. and {Seeley}, D.~H. and {Edmondson}, J.},
        title = "{Alfv{\'e}n Waves in the Solar Corona}",
      journal = {Science},
         year = 2007,
        month = aug,
       volume = {317},
       number = {5842},
        pages = {1192},
          doi = {10.1126/science.1143304},
       adsurl = {https://ui.adsabs.harvard.edu/abs/2007Sci...317.1192T}
}

@ARTICLE{tomczyk2008,
       author = {{Tomczyk}, S. and {Card}, G.~L. and {Darnell}, T. and {Elmore}, D.~F. and {Lull}, R. and {Nelson}, P.~G. and {Streander}, K.~V. and {Burkepile}, J. and {Casini}, R. and {Judge}, P.~G.},
        title = "{An Instrument to Measure Coronal Emission Line Polarization}",
      journal = {Solar Physics},
         year = 2008,
        month = feb,
       volume = {247},
       number = {2},
        pages = {411-428},
          doi = {10.1007/s11207-007-9103-6},
       adsurl = {https://ui.adsabs.harvard.edu/abs/2008SoPh..247..411T}
}

@ARTICLE{Banerjee2021,
       author = {{Banerjee}, D. and {Krishna Prasad}, S. and {Pant}, V. and {McLaughlin}, J.~A. and {Antolin}, P. and {Magyar}, N. and {Ofman}, L. and {Tian}, H. and {Van Doorsselaere}, T. and {De Moortel}, I. and {Wang}, T.~J.},
        title = "{Magnetohydrodynamic Waves in Open Coronal Structures}",
      journal = {Space Science Reviews},
         year = 2021,
        month = oct,
       volume = {217},
       number = {7},
          eid = {76},
        pages = {76},
          doi = {10.1007/s11214-021-00849-0},
archivePrefix = {arXiv},
       eprint = {2012.08802},
 primaryClass = {astro-ph.SR},
       adsurl = {https://ui.adsabs.harvard.edu/abs/2021SSRv..217...76B}
}

@ARTICLE{yang2020Sci,
       author = {{Yang}, Zihao and {Bethge}, Christian and {Tian}, Hui and {Tomczyk}, Steven and {Morton}, Richard and {Del Zanna}, Giulio and {McIntosh}, Scott W. and {Karak}, Bidya Binay and {Gibson}, Sarah and {Samanta}, Tanmoy and {He}, Jiansen and {Chen}, Yajie and {Wang}, Linghua},
        title = "{Global maps of the magnetic field in the solar corona}",
      journal = {Science},
         year = 2020,
        month = aug,
       volume = {369},
       number = {6504},
        pages = {694-697},
          doi = {10.1126/science.abb4462},
archivePrefix = {arXiv},
       eprint = {2008.03136},
 primaryClass = {astro-ph.SR},
       adsurl = {https://ui.adsabs.harvard.edu/abs/2020Sci...369..694Y}
}

@ARTICLE{yang2020,
       author = {{Yang}, ZiHao and {Tian}, Hui and {Tomczyk}, Steven and {Morton}, Richard and {Bai}, XianYong and {Samanta}, Tanmoy and {Chen}, YaJie},
        title = "{Mapping the magnetic field in the solar corona through magnetoseismology}",
      journal = {Science in China E: Technological Sciences},
         year = 2020,
        month = nov,
       volume = {63},
       number = {11},
        pages = {2357-2368},
          doi = {10.1007/s11431-020-1706-9},
archivePrefix = {arXiv},
       eprint = {2008.03146},
 primaryClass = {astro-ph.SR},
       adsurl = {https://ui.adsabs.harvard.edu/abs/2020ScChE..63.2357Y}
}

@ARTICLE{gao2022,
       author = {{Gao}, Yuhang and {Tian}, Hui and {Van Doorsselaere}, Tom and {Chen}, Yajie},
        title = "{Decayless Oscillations in Solar Coronal Bright Points}",
      journal = {the Astrophysical Journal},
         year = 2022,
        month = may,
       volume = {930},
       number = {1},
          eid = {55},
        pages = {55},
          doi = {10.3847/1538-4357/ac62cf},
archivePrefix = {arXiv},
       eprint = {2203.17034},
 primaryClass = {astro-ph.SR},
       adsurl = {https://ui.adsabs.harvard.edu/abs/2022ApJ...930...55G}
}

@ARTICLE{Poletto2015,
       author = {{Poletto}, Giannina},
        title = "{Solar Coronal Plumes}",
      journal = {Living Reviews in Solar Physics},
         year = 2015,
        month = dec,
       volume = {12},
       number = {1},
          eid = {7},
        pages = {7},
          doi = {10.1007/lrsp-2015-7},
       adsurl = {https://ui.adsabs.harvard.edu/abs/2015LRSP...12....7P}
}

@ARTICLE{morton2015,
       author = {{Morton}, R.~J. and {Tomczyk}, S. and {Pinto}, R.},
        title = "{Investigating Alfv{\'e}nic wave propagation in coronal open-field regions}",
      journal = {Nature Communications},
         year = 2015,
        month = jul,
       volume = {6},
          eid = {7813},
        pages = {7813},
          doi = {10.1038/ncomms8813},
       adsurl = {https://ui.adsabs.harvard.edu/abs/2015NatCo...6.7813M}
}

@ARTICLE{lemen2012,
       author = {{Lemen}, James R. and {Title}, Alan M. and {Akin}, David J. and {Boerner}, Paul F. and {Chou}, Catherine and {Drake}, Jerry F. and {Duncan}, Dexter W. and {Edwards}, Christopher G. and {Friedlaender}, Frank M. and {Heyman}, Gary F. and {Hurlburt}, Neal E. and {Katz}, Noah L. and {Kushner}, Gary D. and {Levay}, Michael and {Lindgren}, Russell W. and {Mathur}, Dnyanesh P. and {McFeaters}, Edward L. and {Mitchell}, Sarah and {Rehse}, Roger A. and {Schrijver}, Carolus J. and {Springer}, Larry A. and {Stern}, Robert A. and {Tarbell}, Theodore D. and {Wuelser}, Jean-Pierre and {Wolfson}, C. Jacob and {Yanari}, Carl and {Bookbinder}, Jay A. and {Cheimets}, Peter N. and {Caldwell}, David and {Deluca}, Edward E. and {Gates}, Richard and {Golub}, Leon and {Park}, Sang and {Podgorski}, William A. and {Bush}, Rock I. and {Scherrer}, Philip H. and {Gummin}, Mark A. and {Smith}, Peter and {Auker}, Gary and {Jerram}, Paul and {Pool}, Peter and {Soufli}, Regina and {Windt}, David L. and {Beardsley}, Sarah and {Clapp}, Matthew and {Lang}, James and {Waltham}, Nicholas},
        title = "{The Atmospheric Imaging Assembly (AIA) on the Solar Dynamics Observatory (SDO)}",
      journal = {Solar Physics},
         year = 2012,
        month = jan,
       volume = {275},
       number = {1-2},
        pages = {17-40},
          doi = {10.1007/s11207-011-9776-8},
       adsurl = {https://ui.adsabs.harvard.edu/abs/2012SoPh..275...17L}
}

@ARTICLE{pesnell2012,
       author = {{Pesnell}, W. Dean and {Thompson}, B.~J. and {Chamberlin}, P.~C.},
        title = "{The Solar Dynamics Observatory (SDO)}",
      journal = {Solar Physics},
         year = 2012,
        month = jan,
       volume = {275},
       number = {1-2},
        pages = {3-15},
          doi = {10.1007/s11207-011-9841-3},
       adsurl = {https://ui.adsabs.harvard.edu/abs/2012SoPh..275....3P}
}

@ARTICLE{thurgood2014,
       author = {{Thurgood}, J.~O. and {Morton}, R.~J. and {McLaughlin}, J.~A.},
        title = "{First Direct Measurements of Transverse Waves in Solar Polar Plumes Using SDO/AIA}",
      journal = {the Astrophysical Journal Letters},
         year = 2014,
        month = jul,
       volume = {790},
       number = {1},
          eid = {L2},
        pages = {L2},
          doi = {10.1088/2041-8205/790/1/L2},
archivePrefix = {arXiv},
       eprint = {1406.5348},
 primaryClass = {astro-ph.SR},
       adsurl = {https://ui.adsabs.harvard.edu/abs/2014ApJ...790L...2T}
}

@ARTICLE{weberg2018,
       author = {{Weberg}, Micah J. and {Morton}, Richard J. and {McLaughlin}, James A.},
        title = "{An Automated Algorithm for Identifying and Tracking Transverse Waves in Solar Images}",
      journal = {the Astrophysical Journal},
         year = 2018,
        month = jan,
       volume = {852},
       number = {1},
          eid = {57},
        pages = {57},
          doi = {10.3847/1538-4357/aa9e4a},
archivePrefix = {arXiv},
       eprint = {1807.04842},
 primaryClass = {astro-ph.SR},
       adsurl = {https://ui.adsabs.harvard.edu/abs/2018ApJ...852...57W}
}

@ARTICLE{weberg2020,
       author = {{Weberg}, Micah J. and {Morton}, Richard J. and {McLaughlin}, James A.},
        title = "{Using Transverse Waves to Probe the Plasma Conditions at the Base of the Solar Wind}",
      journal = {the Astrophysical Journal},
         year = 2020,
        month = may,
       volume = {894},
       number = {1},
          eid = {79},
        pages = {79},
          doi = {10.3847/1538-4357/ab7c59},
       adsurl = {https://ui.adsabs.harvard.edu/abs/2020ApJ...894...79W}
}

@ARTICLE{morton2019,
       author = {{Morton}, R.~J. and {Weberg}, M.~J. and {McLaughlin}, J.~A.},
        title = "{A basal contribution from p-modes to the Alfv{\'e}nic wave flux in the Sun's corona}",
      journal = {Nature Astronomy},
         year = 2019,
        month = jan,
       volume = {3},
        pages = {223},
          doi = {10.1038/s41550-018-0668-9},
archivePrefix = {arXiv},
       eprint = {1902.03811},
 primaryClass = {astro-ph.SR},
       adsurl = {https://ui.adsabs.harvard.edu/abs/2019NatAs...3..223M}
}

@ARTICLE{goossens2009,
       author = {{Goossens}, M. and {Terradas}, J. and {Andries}, J. and {Arregui}, I. and {Ballester}, J.~L.},
        title = "{On the nature of kink MHD waves in magnetic flux tubes}",
      journal = {Astronomy and Astrophysics},
         year = 2009,
        month = aug,
       volume = {503},
       number = {1},
        pages = {213-223},
          doi = {10.1051/0004-6361/200912399},
archivePrefix = {arXiv},
       eprint = {0905.0425},
 primaryClass = {astro-ph.SR},
       adsurl = {https://ui.adsabs.harvard.edu/abs/2009A&A...503..213G}
}

@ARTICLE{terradas2010,
       author = {{Terradas}, J. and {Goossens}, M. and {Verth}, G.},
        title = "{Selective spatial damping of propagating kink waves due to resonant absorption}",
      journal = {Astronomy and Astrophysics},
         year = 2010,
        month = dec,
       volume = {524},
          eid = {A23},
        pages = {A23},
          doi = {10.1051/0004-6361/201014845},
archivePrefix = {arXiv},
       eprint = {1004.4468},
 primaryClass = {astro-ph.SR},
       adsurl = {https://ui.adsabs.harvard.edu/abs/2010A&A...524A..23T}
}

@ARTICLE{pagano2019,
       author = {{Pagano}, P. and {De Moortel}, I.},
        title = "{Contribution of observed multi frequency spectrum of Alfv{\'e}n waves to coronal heating}",
      journal = {Astronomy and Astrophysics},
         year = 2019,
        month = mar,
       volume = {623},
          eid = {A37},
        pages = {A37},
          doi = {10.1051/0004-6361/201834158},
archivePrefix = {arXiv},
       eprint = {1901.02310},
 primaryClass = {astro-ph.SR},
       adsurl = {https://ui.adsabs.harvard.edu/abs/2019A&A...623A..37P}
}

@ARTICLE{magyar2017,
       author = {{Magyar}, Norbert and {Van Doorsselaere}, Tom and {Goossens}, Marcel},
        title = "{Generalized phase mixing: Turbulence-like behaviour from unidirectionally propagating MHD waves}",
      journal = {Scientific Reports},
         year = 2017,
        month = nov,
       volume = {7},
          eid = {14820},
        pages = {14820},
          doi = {10.1038/s41598-017-13660-1},
archivePrefix = {arXiv},
       eprint = {1702.02346},
 primaryClass = {astro-ph.SR},
       adsurl = {https://ui.adsabs.harvard.edu/abs/2017NatSR...714820M}
}

@ARTICLE{tvd2020SSRv,
       author = {{Van Doorsselaere}, Tom and {Srivastava}, Abhishek K. and {Antolin}, Patrick and {Magyar}, Norbert and {Vasheghani Farahani}, Soheil and {Tian}, Hui and {Kolotkov}, Dmitrii and {Ofman}, Leon and {Guo}, Mingzhe and {Arregui}, I{\~n}igo and {De Moortel}, Ineke and {Pascoe}, David},
        title = "{Coronal Heating by MHD Waves}",
      journal = {Space Science Reviews},
         year = 2020,
        month = dec,
       volume = {216},
       number = {8},
          eid = {140},
        pages = {140},
          doi = {10.1007/s11214-020-00770-y},
archivePrefix = {arXiv},
       eprint = {2012.01371},
 primaryClass = {astro-ph.SR},
       adsurl = {https://ui.adsabs.harvard.edu/abs/2020SSRv..216..140V}
}

@ARTICLE{Skirvin2024,
       author = {{Skirvin}, S.~J. and {Van Doorsselaere}, T.},
        title = "{Mode conversion and energy flux absorption in the structured solar atmosphere}",
      journal = {Astronomy and Astrophysics},
         year = 2024,
        month = mar,
       volume = {683},
          eid = {A61},
        pages = {A61},
          doi = {10.1051/0004-6361/202348009},
archivePrefix = {arXiv},
       eprint = {2401.02238},
 primaryClass = {astro-ph.SR},
       adsurl = {https://ui.adsabs.harvard.edu/abs/2024A&A...683A..61S}
}

@ARTICLE{shoda2018,
       author = {{Shoda}, Munehito and {Yokoyama}, Takaaki},
        title = "{High-frequency Spicule Oscillations Generated via Mode Conversion}",
      journal = {the Astrophysical Journal},
         year = 2018,
        month = feb,
       volume = {854},
       number = {1},
          eid = {9},
        pages = {9},
          doi = {10.3847/1538-4357/aaa54f},
archivePrefix = {arXiv},
       eprint = {1801.01254},
 primaryClass = {astro-ph.SR},
       adsurl = {https://ui.adsabs.harvard.edu/abs/2018ApJ...854....9S}
}

@ARTICLE{kuniyoshi2024,
       author = {{Kuniyoshi}, Hidetaka and {Shoda}, Munehito and {Morton}, Richard J. and {Yokoyama}, Takaaki},
        title = "{Can the Solar p-modes Contribute to the High-frequency Transverse Oscillations of Spicules?}",
      journal = {the Astrophysical Journal},
         year = 2024,
        month = jan,
       volume = {960},
       number = {2},
          eid = {118},
        pages = {118},
          doi = {10.3847/1538-4357/ad1038},
archivePrefix = {arXiv},
       eprint = {2311.16461},
 primaryClass = {astro-ph.SR},
       adsurl = {https://ui.adsabs.harvard.edu/abs/2024ApJ...960..118K}
}

@ARTICLE{Jess2020,
       author = {{Jess}, David B. and {Snow}, Ben and {Houston}, Scott J. and {Botha}, Gert J.~J. and {Fleck}, Bernhard and {Krishna Prasad}, S. and {Asensio Ramos}, Andr{\'e}s and {Morton}, Richard J. and {Keys}, Peter H. and {Jafarzadeh}, Shahin and {Stangalini}, Marco and {Grant}, Samuel D.~T. and {Christian}, Damian J.},
        title = "{A chromospheric resonance cavity in a sunspot mapped with seismology}",
      journal = {Nature Astronomy},
         year = 2020,
        month = jan,
       volume = {4},
        pages = {220-227},
          doi = {10.1038/s41550-019-0945-2},
       adsurl = {https://ui.adsabs.harvard.edu/abs/2020NatAs...4..220J}
}

@ARTICLE{pant2019,
       author = {{Pant}, Vaibhav and {Magyar}, Norbert and {Van Doorsselaere}, Tom and {Morton}, Richard J.},
        title = "{Investigating {\textquotedblleft}Dark{\textquotedblright} Energy in the Solar Corona Using Forward Modeling of MHD Waves}",
      journal = {the Astrophysical Journal},
         year = 2019,
        month = aug,
       volume = {881},
       number = {2},
          eid = {95},
        pages = {95},
          doi = {10.3847/1538-4357/ab2da3},
archivePrefix = {arXiv},
       eprint = {1906.10941},
 primaryClass = {astro-ph.SR},
       adsurl = {https://ui.adsabs.harvard.edu/abs/2019ApJ...881...95P}
}

@ARTICLE{shi2021forward,
       author = {{Shi}, Mijie and {Van Doorsselaere}, Tom and {Antolin}, Patrick and {Li}, Bo},
        title = "{Forward Modeling of Simulated Transverse Oscillations in Coronal Loops and the Influence of Background Emission}",
      journal = {the Astrophysical Journal},
         year = 2021,
        month = nov,
       volume = {922},
       number = {1},
          eid = {60},
        pages = {60},
          doi = {10.3847/1538-4357/ac2497},
archivePrefix = {arXiv},
       eprint = {2109.02338},
 primaryClass = {astro-ph.SR},
       adsurl = {https://ui.adsabs.harvard.edu/abs/2021ApJ...922...60S}
}

@ARTICLE{guo2023flux,
       author = {{Guo}, Mingzhe and {Gao}, Yuhang and {Van Doorsselaere}, Tom and {Goossens}, Marcel},
        title = "{Estimating the energy flux of transverse waves associated with Kelvin-Helmholtz instability in solar coronal loops}",
      journal = {Astronomy and Astrophysics},
         year = 2023,
        month = aug,
       volume = {676},
          eid = {L7},
        pages = {L7},
          doi = {10.1051/0004-6361/202346816},
       adsurl = {https://ui.adsabs.harvard.edu/abs/2023A&A...676L...7G}
}

@ARTICLE{chen2021,
       author = {{Chen}, Yajie and {Przybylski}, Damien and {Peter}, Hardi and {Tian}, Hui and {Auch{\`e}re}, F. and {Berghmans}, D.},
        title = "{Transient small-scale brightenings in the quiet solar corona: A model for campfires observed with Solar Orbiter}",
      journal = {Astronomy and Astrophysics},
         year = 2021,
        month = dec,
       volume = {656},
          eid = {L7},
        pages = {L7},
          doi = {10.1051/0004-6361/202140638},
archivePrefix = {arXiv},
       eprint = {2104.10940},
 primaryClass = {astro-ph.SR},
       adsurl = {https://ui.adsabs.harvard.edu/abs/2021A&A...656L...7C}
}

@ARTICLE{Gao2024pkw,
       author = {{Gao}, Yuhang and {Van Doorsselaere}, Tom and {Tian}, Hui and {Guo}, Mingzhe and {Karampelas}, Konstantinos},
        title = "{Propagating kink waves in an open coronal magnetic flux tube with gravitational stratification: Magnetohydrodynamic simulation and forward modelling}",
      journal = {Astronomy and Astrophysics},
         year = 2024,
        month = sep,
       volume = {689},
          eid = {A195},
        pages = {A195},
          doi = {10.1051/0004-6361/202450769},
archivePrefix = {arXiv},
       eprint = {2406.19474},
 primaryClass = {astro-ph.SR},
       adsurl = {https://ui.adsabs.harvard.edu/abs/2024A&A...689A.195G}
}

@ARTICLE{bate2022,
       author = {{Bate}, W. and {Jess}, D.~B. and {Nakariakov}, V.~M. and {Grant}, S.~D.~T. and {Jafarzadeh}, S. and {Stangalini}, M. and {Keys}, P.~H. and {Christian}, D.~J. and {Keenan}, F.~P.},
        title = "{High-frequency Waves in Chromospheric Spicules}",
      journal = {the Astrophysical Journal},
         year = 2022,
        month = may,
       volume = {930},
       number = {2},
          eid = {129},
        pages = {129},
          doi = {10.3847/1538-4357/ac5c53},
archivePrefix = {arXiv},
       eprint = {2203.04997},
 primaryClass = {astro-ph.SR},
       adsurl = {https://ui.adsabs.harvard.edu/abs/2022ApJ...930..129B}
}

@ARTICLE{pant2020,
       author = {{Pant}, Vaibhav and {Van Doorsselaere}, Tom},
        title = "{Revisiting the Relation between Nonthermal Line Widths and Transverse MHD Wave Amplitudes}",
      journal = {the Astrophysical Journal},
         year = 2020,
        month = aug,
       volume = {899},
       number = {1},
          eid = {1},
        pages = {1},
          doi = {10.3847/1538-4357/aba429},
archivePrefix = {arXiv},
       eprint = {2007.02836},
 primaryClass = {astro-ph.SR},
       adsurl = {https://ui.adsabs.harvard.edu/abs/2020ApJ...899....1P}
}

@ARTICLE{rochus2020,
       author = {{Rochus}, P. and {Auch{\`e}re}, F. and {Berghmans}, D. and {Harra}, L. and {Schmutz}, W. and {Sch{\"u}hle}, U. and {Addison}, P. and {Appourchaux}, T. and {Aznar Cuadrado}, R. and {Baker}, D. and {Barbay}, J. and {Bates}, D. and {BenMoussa}, A. and {Bergmann}, M. and {Beurthe}, C. and {Borgo}, B. and {Bonte}, K. and {Bouzit}, M. and {Bradley}, L. and {B{\"u}chel}, V. and {Buchlin}, E. and {B{\"u}chner}, J. and {Cab{\'e}}, F. and {Cadiergues}, L. and {Chaigneau}, M. and {Chares}, B. and {Choque Cortez}, C. and {Coker}, P. and {Condamin}, M. and {Coumar}, S. and {Curdt}, W. and {Cutler}, J. and {Davies}, D. and {Davison}, G. and {Defise}, J. -M. and {Del Zanna}, G. and {Delmotte}, F. and {Delouille}, V. and {Dolla}, L. and {Dumesnil}, C. and {D{\"u}rig}, F. and {Enge}, R. and {Fran{\c{c}}ois}, S. and {Fourmond}, J. -J. and {Gillis}, J. -M. and {Giordanengo}, B. and {Gissot}, S. and {Green}, L.~M. and {Guerreiro}, N. and {Guilbaud}, A. and {Gyo}, M. and {Haberreiter}, M. and {Hafiz}, A. and {Hailey}, M. and {Halain}, J. -P. and {Hansotte}, J. and {Hecquet}, C. and {Heerlein}, K. and {Hellin}, M. -L. and {Hemsley}, S. and {Hermans}, A. and {Hervier}, V. and {Hochedez}, J. -F. and {Houbrechts}, Y. and {Ihsan}, K. and {Jacques}, L. and {J{\'e}r{\^o}me}, A. and {Jones}, J. and {Kahle}, M. and {Kennedy}, T. and {Klaproth}, M. and {Kolleck}, M. and {Koller}, S. and {Kotsialos}, E. and {Kraaikamp}, E. and {Langer}, P. and {Lawrenson}, A. and {Le Clech'}, J. -C. and {Lenaerts}, C. and {Liebecq}, S. and {Linder}, D. and {Long}, D.~M. and {Mampaey}, B. and {Markiewicz-Innes}, D. and {Marquet}, B. and {Marsch}, E. and {Matthews}, S. and {Mazy}, E. and {Mazzoli}, A. and {Meining}, S. and {Meltchakov}, E. and {Mercier}, R. and {Meyer}, S. and {Monecke}, M. and {Monfort}, F. and {Morinaud}, G. and {Moron}, F. and {Mountney}, L. and {M{\"u}ller}, R. and {Nicula}, B. and {Parenti}, S. and {Peter}, H. and {Pfiffner}, D. and {Philippon}, A. and {Phillips}, I. and {Plesseria}, J. -Y. and {Pylyser}, E. and {Rabecki}, F. and {Ravet-Krill}, M. -F. and {Rebellato}, J. and {Renotte}, E. and {Rodriguez}, L. and {Roose}, S. and {Rosin}, J. and {Rossi}, L. and {Roth}, P. and {Rouesnel}, F. and {Roulliay}, M. and {Rousseau}, A. and {Ruane}, K. and {Scanlan}, J. and {Schlatter}, P. and {Seaton}, D.~B. and {Silliman}, K. and {Smit}, S. and {Smith}, P.~J. and {Solanki}, S.~K. and {Spescha}, M. and {Spencer}, A. and {Stegen}, K. and {Stockman}, Y. and {Szwec}, N. and {Tamiatto}, C. and {Tandy}, J. and {Teriaca}, L. and {Theobald}, C. and {Tychon}, I. and {van Driel-Gesztelyi}, L. and {Verbeeck}, C. and {Vial}, J. -C. and {Werner}, S. and {West}, M.~J. and {Westwood}, D. and {Wiegelmann}, T. and {Willis}, G. and {Winter}, B. and {Zerr}, A. and {Zhang}, X. and {Zhukov}, A.~N.},
        title = "{The Solar Orbiter EUI instrument: The Extreme Ultraviolet Imager}",
      journal = {Astronomy and Astrophysics},
         year = 2020,
        month = oct,
       volume = {642},
          eid = {A8},
        pages = {A8},
          doi = {10.1051/0004-6361/201936663},
       adsurl = {https://ui.adsabs.harvard.edu/abs/2020A&A...642A...8R}
}

@ARTICLE{muller2020,
       author = {{M{\"u}ller}, D. and {St. Cyr}, O.~C. and {Zouganelis}, I. and {Gilbert}, H.~R. and {Marsden}, R. and {Nieves-Chinchilla}, T. and {Antonucci}, E. and {Auch{\`e}re}, F. and {Berghmans}, D. and {Horbury}, T.~S. and {Howard}, R.~A. and {Krucker}, S. and {Maksimovic}, M. and {Owen}, C.~J. and {Rochus}, P. and {Rodriguez-Pacheco}, J. and {Romoli}, M. and {Solanki}, S.~K. and {Bruno}, R. and {Carlsson}, M. and {Fludra}, A. and {Harra}, L. and {Hassler}, D.~M. and {Livi}, S. and {Louarn}, P. and {Peter}, H. and {Sch{\"u}hle}, U. and {Teriaca}, L. and {del Toro Iniesta}, J.~C. and {Wimmer-Schweingruber}, R.~F. and {Marsch}, E. and {Velli}, M. and {De Groof}, A. and {Walsh}, A. and {Williams}, D.},
        title = "{The Solar Orbiter mission. Science overview}",
      journal = {Astronomy and Astrophysics},
         year = 2020,
        month = oct,
       volume = {642},
          eid = {A1},
        pages = {A1},
          doi = {10.1051/0004-6361/202038467},
archivePrefix = {arXiv},
       eprint = {2009.00861},
 primaryClass = {astro-ph.SR},
       adsurl = {https://ui.adsabs.harvard.edu/abs/2020A&A...642A...1M}
}

@ARTICLE{tian2014,
       author = {{Tian}, H. and {DeLuca}, E.~E. and {Cranmer}, S.~R. and {De Pontieu}, B. and {Peter}, H. and {Mart{\'\i}nez-Sykora}, J. and {Golub}, L. and {McKillop}, S. and {Reeves}, K.~K. and {Miralles}, M.~P. and {McCauley}, P. and {Saar}, S. and {Testa}, P. and {Weber}, M. and {Murphy}, N. and {Lemen}, J. and {Title}, A. and {Boerner}, P. and {Hurlburt}, N. and {Tarbell}, T.~D. and {Wuelser}, J.~P. and {Kleint}, L. and {Kankelborg}, C. and {Jaeggli}, S. and {Carlsson}, M. and {Hansteen}, V. and {McIntosh}, S.~W.},
        title = "{Prevalence of small-scale jets from the networks of the solar transition region and chromosphere}",
      journal = {Science},
         year = 2014,
        month = oct,
       volume = {346},
       number = {6207},
          eid = {1255711},
        pages = {1255711},
          doi = {10.1126/science.1255711},
archivePrefix = {arXiv},
       eprint = {1410.6143},
 primaryClass = {astro-ph.SR},
       adsurl = {https://ui.adsabs.harvard.edu/abs/2014Sci...346A.315T}
}

@ARTICLE{HeSP2020NSR,
       author = {{He}, Shengping and {Wang}, Huijun and {Li}, Fei and {Li}, Hui and {Wang}, Chi},
        title = "{Solar-wind-magnetosphere energy influences the interannual variability of the northern-hemispheric winter climate}",
      journal = {National Science Review},
         year = 2020,
        month = jan,
       volume = {7},
       number = {1},
        pages = {141-148},
          doi = {10.1093/nsr/nwz082},
       adsurl = {https://ui.adsabs.harvard.edu/abs/2020NSRev...7..141H}
}

@ARTICLE{Tarduno2025,
       author = {{Tarduno}, John A. and {Zhou}, Tinghong and {Huang}, Wentao and {Jodder}, Jaganmoy},
        title = "{Earth's magnetic field and its relationship to the origin of life, evolution and planetary habitability}",
      journal = {National Science Review},
         year = 2025,
        month = apr,
       volume = {12},
       number = {5},
          eid = {nwaf082},
        pages = {nwaf082},
          doi = {10.1093/nsr/nwaf082},
       adsurl = {https://ui.adsabs.harvard.edu/abs/2025NSRev..12F..82T}
}

@ARTICLE{Yuan2018,
       author = {{Yuan}, Ding and {Liu}, Wei and {Walsh}, Robert},
        title = "{Investigating Sub-Pixel 45-Second Periodic Wobble in SDO/AIA Data from January to August 2012}",
      journal = {Solar Physics},
         year = 2018,
        month = oct,
       volume = {293},
       number = {10},
          eid = {147},
        pages = {147},
          doi = {10.1007/s11207-018-1368-4},
       adsurl = {https://ui.adsabs.harvard.edu/abs/2018SoPh..293..147Y}
}

@ARTICLE{yang2024,
       author = {{Yang}, Zihao and {Tian}, Hui and {Tomczyk}, Steven and {Liu}, Xianyu and {Gibson}, Sarah and {Morton}, Richard J. and {Downs}, Cooper},
        title = "{Observing the evolution of the Sun's global coronal magnetic field over 8 months}",
      journal = {Science},
         year = 2024,
        month = oct,
       volume = {386},
       number = {6717},
        pages = {76-82},
          doi = {10.1126/science.ado2993},
archivePrefix = {arXiv},
       eprint = {2410.16555},
 primaryClass = {astro-ph.SR},
       adsurl = {https://ui.adsabs.harvard.edu/abs/2024Sci...386...76Y}
}

@ARTICLE{Morton2025origin,
       author = {{Morton}, Richard J. and {Soler}, Roberto},
        title = "{On the Origins of Coronal Alfv{\'e}nic Waves}",
      journal = {the Astrophysical Journal Letters},
         year = 2025,
        month = jun,
       volume = {986},
       number = {1},
          eid = {L6},
        pages = {L6},
          doi = {10.3847/2041-8213/add7da},
archivePrefix = {arXiv},
       eprint = {2505.08636},
 primaryClass = {astro-ph.SR},
       adsurl = {https://ui.adsabs.harvard.edu/abs/2025ApJ...986L...6M}
}

@ARTICLE{tvd2014,
       author = {{Van Doorsselaere}, T. and {Gijsen}, S.~E. and {Andries}, J. and {Verth}, G.},
        title = "{Energy Propagation by Transverse Waves in Multiple Flux Tube Systems Using Filling Factors}",
      journal = {the Astrophysical Journal},
         year = 2014,
        month = nov,
       volume = {795},
       number = {1},
          eid = {18},
        pages = {18},
          doi = {10.1088/0004-637X/795/1/18},
       adsurl = {https://ui.adsabs.harvard.edu/abs/2014ApJ...795...18V}
}

@ARTICLE{YangLP2025,
       author = {{Yang}, Liping and {He}, Jiansen and {Feng}, Xueshang and {Verscharen}, Daniel and {Guo}, Fan and {Li}, Hui and {Tian}, Hui and {Li}, Wenya and {Shen}, Fang and {Hou}, Chuanpeng and {Shi}, Mijie and {Wu}, Honghong and {Xiong}, Ming},
        title = "{Natural Generation of Alfv{\'e}n Waves from Three-dimensional Bursty Interchange Magnetic Reconnection in the Solar Corona}",
      journal = {the Astrophysical Journal Letters},
         year = 2025,
        month = mar,
       volume = {982},
       number = {1},
          eid = {L25},
        pages = {L25},
          doi = {10.3847/2041-8213/adb8ce},
       adsurl = {https://ui.adsabs.harvard.edu/abs/2025ApJ...982L..25Y}
}

@ARTICLE{Lynch2014,
       author = {{Lynch}, B.~J. and {Edmondson}, J.~K. and {Li}, Y.},
        title = "{Interchange Reconnection Alfv{\'e}n Wave Generation}",
      journal = {Solar Physics},
         year = 2014,
        month = aug,
       volume = {289},
       number = {8},
        pages = {3043-3058},
          doi = {10.1007/s11207-014-0506-x},
archivePrefix = {arXiv},
       eprint = {1401.7965},
 primaryClass = {astro-ph.SR},
       adsurl = {https://ui.adsabs.harvard.edu/abs/2014SoPh..289.3043L}
}

@ARTICLE{Shrivastav2024,
       author = {{Shrivastav}, Arpit Kumar and {Pant}, Vaibhav and {Berghmans}, David and {Zhukov}, Andrei N. and {Van Doorsselaere}, Tom and {Petrova}, Elena and {Banerjee}, Dipankar and {Lim}, Daye and {Verbeeck}, Cis},
        title = "{Statistical investigation of decayless oscillations in small-scale coronal loops observed by Solar Orbiter/EUI}",
      journal = {Astronomy and Astrophysics},
         year = 2024,
        month = may,
       volume = {685},
          eid = {A36},
        pages = {A36},
          doi = {10.1051/0004-6361/202346670},
archivePrefix = {arXiv},
       eprint = {2304.13554},
 primaryClass = {astro-ph.SR},
       adsurl = {https://ui.adsabs.harvard.edu/abs/2024A&A...685A..36S}
}

@ARTICLE{Lim2024,
       author = {{Lim}, Daye and {Van Doorsselaere}, Tom and {Berghmans}, David and {Petrova}, Elena},
        title = "{Characteristics and energy flux distributions of decayless transverse oscillations depending on coronal regions}",
      journal = {Astronomy and Astrophysics},
         year = 2024,
        month = sep,
       volume = {689},
          eid = {A16},
        pages = {A16},
          doi = {10.1051/0004-6361/202450433},
archivePrefix = {arXiv},
       eprint = {2406.06368},
 primaryClass = {astro-ph.SR},
       adsurl = {https://ui.adsabs.harvard.edu/abs/2024A&A...689A..16L}
}

@ARTICLE{tvd2020ApJ,
       author = {{Van Doorsselaere}, Tom and {Li}, Bo and {Goossens}, Marcel and {Hnat}, Bogdan and {Magyar}, Norbert},
        title = "{Wave Pressure and Energy Cascade Rate of Kink Waves Computed with Els{\"a}sser Variables}",
      journal = {the Astrophysical Journal},
         year = 2020,
        month = aug,
       volume = {899},
       number = {2},
          eid = {100},
        pages = {100},
          doi = {10.3847/1538-4357/aba0b8},
archivePrefix = {arXiv},
       eprint = {2007.15411},
 primaryClass = {astro-ph.SR},
       adsurl = {https://ui.adsabs.harvard.edu/abs/2020ApJ...899..100V}
}

@ARTICLE{SPO2025,
       author = {{Deng}, Yuanyong and {Tian}, Hui and {Jiang}, Jie and {Yang}, Shuhong and {Li}, Hao and {Cameron}, Robert and {Gizon}, Laurent and {Harra}, Louise and {Wimmer-Schweingruber}, Robert F. and {Auch{\`e}re}, Fr{\'e}d{\'e}ric and {Bai}, Xianyong and {Bellot}, Rubio Luis and {Chen}, Linjie and {Chen}, Pengfei and {Chitta}, Lakshmi Pradeep and {Davies}, Jackie and {Favata}, Fabio and {Feng}, Li and {Feng}, Xueshang and {Gan}, Weiqun and {Hassler}, Don and {He}, Jiansen and {Hou}, Junfeng and {Hou}, Zhenyong and {Jin}, Chunlan and {Li}, Wenya and {Lin}, Jiaben and {Nandy}, Dibyendu and {Pant}, Vaibhav and {Romoli}, Marco and {Sakao}, Taro and {Krishna Prasad}, Sayamanthula and {Shen}, Fang and {Su}, Yang and {Toriumi}, Shin and {Tripathi}, Durgesh and {Wang}, Linghua and {Wang}, Jingjing and {Xia}, Lidong and {Xiong}, Ming and {Yan}, Yihua and {Yang}, Liping and {Yang}, Shangbin and {Zhang}, Mei and {Zhou}, Guiping and {Zhu}, Xiaoshuai and {Wang}, Jingxiu and {Wang}, Chi},
        title = "{Probing Solar Polar Regions}",
      journal = {Chinese Journal of Space Science},
         year = 2025,
        month = jul,
       volume = {45},
       number = {4},
        pages = {913-942},
          doi = {10.11728/cjss2025.04.2025-0054},
archivePrefix = {arXiv},
       eprint = {2506.20502},
 primaryClass = {astro-ph.SR},
       adsurl = {https://ui.adsabs.harvard.edu/abs/2025ChJSS..45..913D}
}

@ARTICLE{He2009,
       author = {{He}, J.-S. and {Tu}, C.-Y. and {Marsch}, E. and {Guo}, L.-J. and {Yao}, S. and {Tian}, H.},
        title = "{Upward propagating high-frequency Alfv{\'e}n waves as identified from dynamic wave-like spicules observed by SOT on Hinode}",
      journal = {Astronomy and Astrophysics},
         year = 2009,
        month = apr,
       volume = {497},
       number = {2},
        pages = {525-535},
          doi = {10.1051/0004-6361/200810777},
       adsurl = {https://ui.adsabs.harvard.edu/abs/2009A&A...497..525H}
}

@ARTICLE{Khomenko2012,
       author = {{Khomenko}, E. and {Collados}, M.},
        title = "{Heating of the Magnetized Solar Chromosphere by Partial Ionization Effects}",
      journal = {the Astrophysical Journal},
         year = 2012,
        month = mar,
       volume = {747},
       number = {2},
          eid = {87},
        pages = {87},
          doi = {10.1088/0004-637X/747/2/87},
archivePrefix = {arXiv},
       eprint = {1112.3374},
 primaryClass = {astro-ph.SR},
       adsurl = {https://ui.adsabs.harvard.edu/abs/2012ApJ...747...87K}
}

@ARTICLE{Miriyala2025,
       author = {{Miriyala}, Hemanthi and {Morton}, Richard J. and {Khomenko}, Elena and {Antolin}, Patrick and {Botha}, Gert J.~J.},
        title = "{The Coronal Power Spectrum from MHD Mode Conversion above Sunspots}",
      journal = {the Astrophysical Journal},
         year = 2025,
        month = feb,
       volume = {979},
       number = {2},
          eid = {236},
        pages = {236},
          doi = {10.3847/1538-4357/ada26f},
archivePrefix = {arXiv},
       eprint = {2501.03094},
 primaryClass = {astro-ph.SR},
       adsurl = {https://ui.adsabs.harvard.edu/abs/2025ApJ...979..236M}
}

@ARTICLE{Fehlmann2023,
       author = {{Fehlmann}, Andr{\'e} and {Kuhn}, Jeffrey R. and {Schad}, Thomas A. and {Scholl}, Isabelle F. and {Williams}, Rebecca and {Agdinaoay}, Rodell and {Berst}, D. Christopher and {Craig}, Simon C. and {Giebink}, Cynthia and {Goodrich}, Bret and {Hnat}, Kirby and {James}, Don and {Lockhart}, Charles and {Mickey}, Donald L. and {Oswald}, Daniel and {Puentes}, Myles M. and {Schickling}, Richard and {de Vanssay}, Jean-Benoit and {Warmbier}, Eric A.},
        title = "{The Daniel K. Inouye Solar Telescope (DKIST) Cryogenic Near-Infrared Spectro-Polarimeter}",
      journal = {Solar Physics},
         year = 2023,
        month = jan,
       volume = {298},
       number = {1},
          eid = {5},
        pages = {5},
          doi = {10.1007/s11207-022-02098-y},
       adsurl = {https://ui.adsabs.harvard.edu/abs/2023SoPh..298....5F}
}

@ARTICLE{DKIST2020,
       author = {{Rimmele}, Thomas R. and {Warner}, Mark and {Keil}, Stephen L. and {Goode}, Philip R. and {Kn{\"o}lker}, Michael and {Kuhn}, Jeffrey R. and {Rosner}, Robert R. and {McMullin}, Joseph P. and {Casini}, Roberto and {Lin}, Haosheng and {W{\"o}ger}, Friedrich and {von der L{\"u}he}, Oskar and {Tritschler}, Alexandra and {Davey}, Alisdair and {de Wijn}, Alfred and {Elmore}, David F. and {Fehlmann}, Andr{\'e} and {Harrington}, David M. and {Jaeggli}, Sarah A. and {Rast}, Mark P. and {Schad}, Thomas A. and {Schmidt}, Wolfgang and {Mathioudakis}, Mihalis and {Mickey}, Donald L. and {Anan}, Tetsu and {Beck}, Christian and {Marshall}, Heather K. and {Jeffers}, Paul F. and {Oschmann}, Jacobus M. and {Beard}, Andrew and {Berst}, David C. and {Cowan}, Bruce A. and {Craig}, Simon C. and {Cross}, Eric and {Cummings}, Bryan K. and {Donnelly}, Colleen and {de Vanssay}, Jean-Benoit and {Eigenbrot}, Arthur D. and {Ferayorni}, Andrew and {Foster}, Christopher and {Galapon}, Chriselle Ann and {Gedrites}, Christopher and {Gonzales}, Kerry and {Goodrich}, Bret D. and {Gregory}, Brian S. and {Guzman}, Stephanie S. and {Guzzo}, Stephen and {Hegwer}, Steve and {Hubbard}, Robert P. and {Hubbard}, John R. and {Johansson}, Erik M. and {Johnson}, Luke C. and {Liang}, Chen and {Liang}, Mary and {McQuillen}, Isaac and {Mayer}, Christopher and {Newman}, Karl and {Onodera}, Brialyn and {Phelps}, LeEllen and {Puentes}, Myles M. and {Richards}, Christopher and {Rimmele}, Lukas M. and {Sekulic}, Predrag and {Shimko}, Stephan R. and {Simison}, Brett E. and {Smith}, Brett and {Starman}, Erik and {Sueoka}, Stacey R. and {Summers}, Richard T. and {Szabo}, Aimee and {Szabo}, Louis and {Wampler}, Stephen B. and {Williams}, Timothy R. and {White}, Charles},
        title = "{The Daniel K. Inouye Solar Telescope - Observatory Overview}",
      journal = {Solar Physics},
         year = 2020,
        month = dec,
       volume = {295},
       number = {12},
          eid = {172},
        pages = {172},
          doi = {10.1007/s11207-020-01736-7},
       adsurl = {https://ui.adsabs.harvard.edu/abs/2020SoPh..295..172R}
}

@ARTICLE{Baweja2025,
       author = {{Baweja}, Upasna and {Pant}, Vaibhav and {Krishna Prasad}, S. and {Shrivastav}, Arpit Kumar and {Van Doorsselaere}, Tom and {Narang}, Nancy and {Verbeeck}, Cis and {Khan}, M. Saleem and {Berghmans}, David},
        title = "{Coexistence of Longitudinal and Transverse Oscillations in Polar Plumes Observed with Solar Orbiter/Extreme Ultraviolet Imager}",
      journal = {the Astrophysical Journal Letters},
         year = 2025,
        month = oct,
       volume = {991},
       number = {2},
          eid = {L45},
        pages = {L45},
          doi = {10.3847/2041-8213/ae071e},
archivePrefix = {arXiv},
       eprint = {2509.07796},
 primaryClass = {astro-ph.SR},
       adsurl = {https://ui.adsabs.harvard.edu/abs/2025ApJ...991L..45B}
}

@ARTICLE{DeForest2018,
       author = {{DeForest}, C.~E. and {Howard}, R.~A. and {Velli}, M. and {Viall}, N. and {Vourlidas}, A.},
        title = "{The Highly Structured Outer Solar Corona}",
      journal = {the Astrophysical Journal},
         year = 2018,
        month = jul,
       volume = {862},
       number = {1},
          eid = {18},
        pages = {18},
          doi = {10.3847/1538-4357/aac8e3},
       adsurl = {https://ui.adsabs.harvard.edu/abs/2018ApJ...862...18D}
}

@ARTICLE{Goossens2011,
       author = {{Goossens}, Marcel and {Erd{\'e}lyi}, Robert and {Ruderman}, Michael S.},
        title = "{Resonant MHD Waves in the Solar Atmosphere}",
      journal = {Space Science Reviews},
         year = 2011,
        month = jul,
       volume = {158},
       number = {2-4},
        pages = {289-338},
          doi = {10.1007/s11214-010-9702-7},
       adsurl = {https://ui.adsabs.harvard.edu/abs/2011SSRv..158..289G}
}

@ARTICLE{Samanta2015,
       author = {{Samanta}, Tanmoy and {Pant}, Vaibhav and {Banerjee}, Dipankar},
        title = "{Propagating Disturbances in the Solar Corona and Spicular Connection}",
      journal = {the Astrophysical Journal Letters},
         year = 2015,
        month = dec,
       volume = {815},
       number = {1},
          eid = {L16},
        pages = {L16},
          doi = {10.1088/2041-8205/815/1/L16},
archivePrefix = {arXiv},
       eprint = {1511.07354},
 primaryClass = {astro-ph.SR},
       adsurl = {https://ui.adsabs.harvard.edu/abs/2015ApJ...815L..16S}
}

@ARTICLE{Jiao2015,
       author = {{Jiao}, Fangran and {Xia}, Lidong and {Li}, Bo and {Huang}, Zhenghua and {Li}, Xing and {Chandrashekhar}, Kalugodu and {Mou}, Chaozhou and {Fu}, Hui},
        title = "{Sources of Quasi-periodic Propagating Disturbances above a Solar Polar Coronal Hole}",
      journal = {the Astrophysical Journal Letters},
         year = 2015,
        month = aug,
       volume = {809},
       number = {1},
          eid = {L17},
        pages = {L17},
          doi = {10.1088/2041-8205/809/1/L17},
archivePrefix = {arXiv},
       eprint = {1507.08440},
 primaryClass = {astro-ph.SR},
       adsurl = {https://ui.adsabs.harvard.edu/abs/2015ApJ...809L..17J}
}

@ARTICLE{Morton2025NA,
author = {{Morton}, R.~J. and {Gao}, Y. and {Tajfirouze}, E. and {Tian}, H. and {Van Doorsselaere}, T. and {Schad}, T.~A.},
title = "{Evidence for small-scale torsional Alfv{\'e}n waves in the solar corona}",
journal = {Nature Astronomy},
year = 2026,
month = jan,
volume = {10},
pages = {42-53},
doi = {10.1038/s41550-025-02690-9},
adsurl = {https://ui.adsabs.harvard.edu/abs/2026NatAs..10...42M}
}

@ARTICLE{Madjarska2024,
	author = {{Madjarska}, Maria S. and {Wiegelmann}, Thomas and {D{\'e}moulin}, Pascal and {Galsgaard}, Klaus},
	title = "{Coronal magnetic field and emission properties of small-scale bright and faint loops in the quiet Sun}",
	journal = {Astronomy and Astrophysics},
	year = 2024,
	month = oct,
	volume = {690},
	eid = {A242},
	pages = {A242},
	doi = {10.1051/0004-6361/202450343},
	archivePrefix = {arXiv},
	eprint = {2407.09769},
	primaryClass = {astro-ph.SR},
	adsurl = {https://ui.adsabs.harvard.edu/abs/2024A&A...690A.242M}
}

@ARTICLE{Morton2025cycle,
	author = {{Morton}, R.~J. and {Weberg}, M.~J. and {Balodhi}, N. and {McLaughlin}, J.~A.},
	title = "{Estimating the Poynting Flux of Alfv{\'e}nic Waves in Polar Coronal Holes across Solar Cycle 24}",
	journal = {the Astrophysical Journal},
	year = 2025,
	month = may,
	volume = {985},
	number = {1},
	eid = {13},
	pages = {13},
	doi = {10.3847/1538-4357/adc568},
	archivePrefix = {arXiv},
	eprint = {2501.13673},
	primaryClass = {astro-ph.SR},
	adsurl = {https://ui.adsabs.harvard.edu/abs/2025ApJ...985...13M}
}

@ARTICLE{Long2023,
	author = {{Long}, David M. and {Chitta}, Lakshmi Pradeep and {Baker}, Deborah and {Hannah}, Iain G. and {Ngampoopun}, Nawin and {Berghmans}, David and {Zhukov}, Andrei N. and {Teriaca}, Luca},
	title = "{Multistage Reconnection Powering a Solar Coronal Jet}",
	journal = {the Astrophysical Journal},
	year = 2023,
	month = feb,
	volume = {944},
	number = {1},
	eid = {19},
	pages = {19},
	doi = {10.3847/1538-4357/acb0c9},
	archivePrefix = {arXiv},
	eprint = {2301.02034},
	primaryClass = {astro-ph.SR},
	adsurl = {https://ui.adsabs.harvard.edu/abs/2023ApJ...944...19L}
}

@ARTICLE{HuangZG2024,
	author = {{Huang}, Zhenguang and {T{\'o}th}, G{\'a}bor and {Sachdeva}, Nishtha and {van der Holst}, Bart},
	title = "{Solar Wind Driven from GONG Magnetograms in the Last Solar Cycle}",
	journal = {the Astrophysical Journal},
	year = 2024,
	month = apr,
	volume = {965},
	number = {1},
	eid = {1},
	pages = {1},
	doi = {10.3847/1538-4357/ad32ca},
	archivePrefix = {arXiv},
	eprint = {2403.01656},
	primaryClass = {astro-ph.SR},
	adsurl = {https://ui.adsabs.harvard.edu/abs/2024ApJ...965....1H}
}

@ARTICLE{Huang2021,
	author = {{Huang}, Zhenghua and {Zhang}, Quanhao and {Xia}, Lidong and {Feng}, Li and {Fu}, Hui and {Liu}, Weixin and {Sun}, Mingzhe and {Qi}, Youqian and {Liu}, Dayang and {Zhang}, Qingmin and {Li}, Bo},
	title = "{Population of Bright Plume Threads in Solar Polar Coronal Holes}",
	journal = {Solar Physics},
	year = 2021,
	month = jan,
	volume = {296},
	number = {1},
	eid = {22},
	pages = {22},
	doi = {10.1007/s11207-021-01773-w},
	archivePrefix = {arXiv},
	eprint = {2101.03768},
	primaryClass = {astro-ph.SR},
	adsurl = {https://ui.adsabs.harvard.edu/abs/2021SoPh..296...22H}
}

@ARTICLE{Moore2023,
	author = {{Moore}, Ronald L. and {Tiwari}, Sanjiv K. and {Panesar}, Navdeep K. and {Sterling}, Alphonse C.},
	title = "{Prospective Implications of Extreme-ultraviolet Coronal Plumes for Magnetic-network Genesis of Coronal Heating, Coronal-hole Solar Wind, and Solar-wind Magnetic Field Switchbacks}",
	journal = {the Astrophysical Journal Letters},
	year = 2023,
	month = mar,
	volume = {945},
	number = {1},
	eid = {L16},
	pages = {L16},
	doi = {10.3847/2041-8213/acbe46},
	archivePrefix = {arXiv},
	eprint = {2303.00097},
	primaryClass = {astro-ph.SR},
	adsurl = {https://ui.adsabs.harvard.edu/abs/2023ApJ...945L..16M}
}

@ARTICLE{Alzate2025,
	author = {{Alzate}, Nathalia and {Di Matteo}, Simone and {Higginson}, Aleida},
	title = "{Coronal Cells in Coronal Holes: Systematic Analysis and Implications for Coronal Evolution}",
	journal = {the Astrophysical Journal},
	year = 2025,
	month = sep,
	volume = {991},
	number = {1},
	eid = {55},
	pages = {55},
	doi = {10.3847/1538-4357/adf7a9},
	archivePrefix = {arXiv},
	eprint = {2508.03607},
	primaryClass = {astro-ph.SR},
	adsurl = {https://ui.adsabs.harvard.edu/abs/2025ApJ...991...55A}
}

@ARTICLE{Tu2005,
	author = {{Tu}, Chuan-Yi and {Zhou}, Cheng and {Marsch}, Eckart and {Xia}, Li-Dong and {Zhao}, Liang and {Wang}, Jing-Xiu and {Wilhelm}, Klaus},
	title = "{Solar Wind Origin in Coronal Funnels}",
	journal = {Science},
	year = 2005,
	month = apr,
	volume = {308},
	number = {5721},
	pages = {519-523},
	doi = {10.1126/science.1109447},
	adsurl = {https://ui.adsabs.harvard.edu/abs/2005Sci...308..519T}
}

@ARTICLE{Marsch1997,
       author = {{Marsch}, E. and {Tu}, C.-Y.},
        title = "{The effects of high-frequency Alfven waves on coronal heating and solar wind acceleration.}",
      journal = {Astronomy and Astrophysics},
         year = 1997,
        month = mar,
       volume = {319},
        pages = {L17-L20},
       adsurl = {https://ui.adsabs.harvard.edu/abs/1997A&A...319L..17M}
}

@ARTICLE{Liu2014,
       author = {{Liu}, Zi-Xu and {He}, Jian-Sen and {Yan}, Li-Mei},
        title = "{Observations of counter-propagating Alfv{\'e}nic and compressive fluctuations in the chromosphere}",
      journal = {Research in Astronomy and Astrophysics},
         year = 2014,
        month = mar,
       volume = {14},
       number = {3},
          eid = {299-310},
        pages = {299-310},
          doi = {10.1088/1674-4527/14/3/004},
       adsurl = {https://ui.adsabs.harvard.edu/abs/2014RAA....14..299L}
}

\end{document}